\documentclass[12pt]{article}
\usepackage{epsfig,cite}
\usepackage {amsmath}
\usepackage{amssymb}
\include{epsf}
\newcount\timecount
\newcount\hours \newcount\minutes  \newcount\temp \newcount\pmhours
\hours = \time
\divide\hours by 60
\temp = \hours
\multiply\temp by 60
\minutes = \time
\advance\minutes by -\temp
\def\hour{\the\hours}
\def\minute{\ifnum\minutes<10 0\the\minutes
            \else\the\minutes\fi}
\def\clock{
\ifnum\hours=0 12:\minute\ AM
\else\ifnum\hours<12 \hour:\minute\ AM
      \else\ifnum\hours=12 12:\minute\ PM
            \else\ifnum\hours>12
                 \pmhours=\hours
                 \advance\pmhours by -12
                 \the\pmhours:\minute\ PM
                 \fi
            \fi
      \fi
\fi
}

\def\monthname{\relax\ifcase\month 0/\or January\or February\or
   March\or April\or May\or June\or July\or August\or September\or
   October\or November\or December\else\number\month/\fi}

\def\bold#1{\setbox0=\hbox{$#1$}%
     \kern-.025em\copy0\kern-\wd0
     \kern.05em\copy0\kern-\wd0
     \kern-.025em\raise.0433em\box0 }

\def\beq{\begin{equation}}
\def\eeq{\end{equation}}

\def\ga{\mathrel{\raise.3ex\hbox{$>$\kern-.75em\lower1ex\hbox{$\sim$}}}}
\def\la{\mathrel{\raise.3ex\hbox{$<$\kern-.75em\lower1ex\hbox{$\sim$}}}}
\def\gev{{\rm \, Ge\kern-0.125em V}}
\def\tev{{\rm \, Te\kern-0.125em V}}
\def\gyr{{\rm \, G\kern-0.125em yr}}

\def\tbt{\tan \beta}

\def\gappeq{\mathrel{\rlap {\raise.5ex\hbox{$>$}}
{\lower.5ex\hbox{$\sim$}}}}
\def\lappeq{\mathrel{\rlap{\raise.5ex\hbox{$<$}}
{\lower.5ex\hbox{$\sim$}}}}
\def\Toprel#1\over#2{\mathrel{\mathop{#2}\limits^{#1}}}

\def\m12{m_{1\!/2}}

\def\five{{\bf 5}}
\def\fivebar{\overline{\bf 5}}
\def\ten{{\bf 10}}
\def\tenbar{\overline{\bf 10}}

\def\bea{\begin{eqnarray}}
\def\eea{\end{eqnarray}}

\def\beq{\begin{equation}}
\def\eeq{\end{equation}}

\begin{document}

\begin{titlepage}
\pagestyle{empty}
\baselineskip=21pt
\rightline{UMN--TH--3413/14, FTPI--MINN--14/42}
\vspace{0.2cm}
\begin{center}
{\large {\bf Light Higgsinos in Pure Gravity Mediation }}
\end{center}
\vspace{0.5cm}
\begin{center}
{\bf Jason L. Evans}$^{1}$,
{\bf Masahiro Ibe}$^{2,3}$ {\bf Keith~A.~Olive}$^{1}$
and {\bf Tsutomu T. Yanagida}$^{3}$\\
\vskip 0.2in
{\small {\it
$^1${William I. Fine Theoretical Physics Institute, School of Physics and Astronomy},\\
{University of Minnesota, Minneapolis, MN 55455,\,USA}\\
$^2${ ICRR, University of Tokyo, Kashiwa 277-8582, Japan}\\
$^3${Kavli IPMU, TODIAS, University of Tokyo, Kashiwa 277-8583, Japan}\\
}}
\vspace{1cm}
\vspace{1cm}
{\bf Abstract}
\end{center}
\baselineskip=18pt \noindent
{\small
Pure gravity mediation (PGM), with two free parameters, is a minimalistic approach to supergravity
models, yet is capable of incorporating radiative electroweak symmetry breaking, a Higgs mass
in agreement with the experimental measurement, without violating any phenomenological constraints.
The model may also contain a viable dark matter candidate in the form of a wino.
Here, we extend the minimal model by allowing the $\mu$-term to be a free parameter
(equivalent to allowing the two Higgs soft masses, $m_1$ and $m_2$,  to differ from other scalar
masses) which are set by the gravitino mass. In particular, we examine the region of
parameter space where $\mu \ll m_{3/2}$ in which case, the Higgsino becomes the
lightest supersymmetric particle (LSP) and a dark matter candidate. We also consider
a generalization of PGM which incorporates a Peccei-Quinn symmetry which
determines the $\mu$-term dynamically. In this case, we show that the dark matter
may either be in the form of an axion and/or a neutralino, and that the LSP may
be either a wino, bino, or Higgsino.
}


\vfill

\end{titlepage}

\section{Introduction}
Because of the rather strong constraints on sparticles masses from the LHC \cite{lhc}, mass spectra with larger sfermion masses have become more relevant \cite{highmass,nuhm2high}.  Even simple models like the constrained minimal supersymmetric standard model (CMSSM) \cite{cmssm} cannot avoid sfermion masses larger than about $1$ TeV.  Although large sfermion masses strain one of the original motivations for supersymmetry (SUSY), there are benefits to considering models with heavier sfermions. Once we give up the notion that the sfermions and gauginos must have similar masses, models such as the CMSSM can be further simplified. For example, models such as pure gravity mediation (PGM)  \cite{pgm,pgm2,ArkaniHamed:2012gw,eioy,eioy2,eo} can be formulated in terms of a single parameter. This reduction in parameters is even more tantalizing because it comes with a solution of Polonyi problem \cite{polprob,polprob2}, which has plagued many models based on supergravity (SUGRA) \cite{sol1,sol2}. In addition, if these theories
are expected to originate at some UV scale, they should also account for electroweak symmetry breaking
 (EWSB) and this is most elegantly done radiatively \cite{ewsb}.

The minimal model of pure gravity mediation (PGM) assumes a flat K\"ahler potential which gives universal sfermion masses equal to the gravitino mass, $m_{3/2}$.  The gaugino masses are generated from anomalies and so are loop suppressed relative to the sfermion masses \cite{anom}.  The $B$ parameter is set by SUGRA, $B_0=A_0-m_{3/2}$, and it is assumed that there is a tree-level Higgs bilinear mass term $\mu$. In PGM models, the $A$-terms are generated by anomalies and are small enough that we can ignore them. Thus, effectively, we have $B_0=-m_{3/2}$.  Using the measured value of $M_Z$, $\mu$ and the ratio of the Higgs vacuum expectation values, $\tan\beta$, are determined by radiative electroweak symmetry breaking leaving the single free parameter as $m_{3/2}$.  Because the EWSB parameters are overly constrained, this simplest of models fails to achieve radiative electroweak symmetry breaking.

There are several ways to relax the constraints coming from radiative electroweak symmetry breaking.  The simplest choice is to add a Giudice-Masiero (GM) term \cite{gm,ikyy,dmmo} to the K\"ahler potential, which was considered in \cite{eioy}. The GM term in the K\"ahler potential combined with a tree level $\mu_0$ in the superpotential allows the total $\mu$ and $B$ to be completely independent parameters and radiative electroweak symmetry breaking can occur. In practice, one can trade the GM term for $\tan \beta$ and use the EWSB conditions to determine the GM coupling. This leaves two free parameters, $m_{3/2}$ and $\tan\beta$. Because the Higgs mass has now been measured \cite{lhch}, only one of the two parameters can be adjusted independently. As it turns out, even in these two-parameter models of PGM, $\tan\beta$ is highly restricted and achieving a large enough Higgs mass is non-trivial.  The constraints on $\tan\beta$ can be further relaxed if the Higgs soft masses are allowed to deviate from their universal value \cite{eioy2}.  In this case, $\tan\beta$ becomes relatively unconstrained and a larger Higgs mass is achieved relatively easily.

In each of these variants of PGM, the lightest supersymmetric particle (LSP) is usually very wino-like.
To get a thermal relic wino dark matter candidate, $m_{\tilde W}\simeq 3$ TeV, $m_{3/2}$ needs to be about $450$ TeV. With $m_{3/2}$ this large, detection of the particle spectrum implied by these models at the LHC is very unlikely.
Even if we abandon hope for discovery at the LHC, the large annihilation cross section
of wino dark matter is in tension with gamma-ray observations of the Galactic center
in the Fermi-LAT and the H.E.S.S. telescope \cite{wino},
although there is still some ambiguity in the dark matter profile at the Galactic center.
This has put the natural dark matter candidate of PGM under pressure and motivates alternative candidates for dark matter.

There are several ways to get additional dark matter candidates.
One possibility is to add vector-like multiplets \cite{hiy,eo}. For example, by adding either
a $\ten, \tenbar$ pair, or pairs of $\five$ and $\fivebar$ multiplets, one affects the beta functions for the
standard model gauge couplings. As a result the commonly adopted ratios of gaugino masses
at the GUT scale
in anomaly mediated models, $M_1:M_2:M_3 = 11:1:-3$, is altered by adding
$\left(N_{5+\bar 5} + 3N_{10+\bar 10}\right)$ to each term (remembering a factor of 5/3 for $M_1$).
At low energies, it becomes possible for the bino (or gluino) to become the LSP.

A simpler possibility is realized if the Higgs bilinear mass $\mu$ is chosen to be small enough so that the Higgsino is the LSP. We will explore this option in the context of several simple models. In PGM with
radiative EWSB, we consider $m_{3/2}$ and $\tan \beta$ as free input parameters, and use the
EWSB conditions to solve for the GM coupling, $c_H$ and tree-level $\mu$-term, $\mu_0$, while
maintaining full scalar mass universality. The effective $\mu$-term is given by $\mu_0 + c_H m_{3/2}$
and the effective $B$-term is given by $B\mu = B_0 \mu_0 + 2 c_H m_{3/2}^2$. If we
choose $\mu$ as a third free parameter we must allow the Higgs soft masses to differ from $m_{3/2}$
though in this case, we may keep $m_1 = m_2$ as in NUHM1 models \cite{nuhm1}.
In this case, it is possible to get radiative EWSB and have a Higgsino LSP which is a good dark matter candidate. Although, this does fix the dark matter problem it does present other problems. A GM term forces the operator $H_uH_d$ to have an $R$ charge of zero. Where on the other hand, a supersymmetric bilinear term in the superpotential prefers $H_uH_d$ to have $R$ charge of two. Although, it is possible to generate a Higgs bilinear mass from non-renormalizable operators, which can meet the $R$ charge constraints and keep a GM term,  without a model to justify these non-renormalizable operators it is less appealing.

Instead,  we may also consider the
presence of a tree-level $\mu$-term in the superpotential with no additional GM coupling in the K\"ahler potential.  Supergravity effects will then generate a $B$ term given by $B_0=-m_{3/2}$. In this case,  we must allow the two Higgs soft masses to independently differ from $m_{3/2}$ in order to the satisfy the electroweak symmetry breaking conditions, much like NUHM2 models \cite{nuhm2}. As a related option we can assume no tree-level $\mu_0$ in the superpotential and assume that both $\mu$ and $B$ are completely determined by a GM term, $c_H$, in the K\"ahler potential. In this case,  $\mu=c_Hm_{3/2}$ and $B=2m_{3/2}$. As we will see, neither of these models will help us obtain Higgsino dark matter.

A possibly more appealing solution, is to consider additional candidates for dark matter other than the SUSY LSP. One good candidate for this is the axion. If we consider a DFSZ\cite{DFSZ} axion, the Higgses are charged under the PQ symmetry. The Higgs fields can then couple to the PQ breaking fields which will generate independent $\mu$ and $B$ terms. In \cite{eioy4}, it was shown that this mechanism combined with a vanishing GM term allows for radiative electroweak symmetry breaking. It can also produce a viable axion dark matter candidate if the relatively high scale for PQ breaking is generated before inflation and if one employs some novel techniques to solve the problem of  isocurvature perturbations. Another possible way to solve the isocurvature problem is to consider a lower PQ breaking scale with PQ breaking after inflation. As long as $N_{DW}=1$, PQ breaking after inflation is not problematic. Since the the PQ breaking field are still coupled to $H_uH_d$, they will again generate independent $B$ and $\mu$ terms. However, these models will require additional $SU(3)$ charged fields to be coupled to the PQ breaking fields so that we get $N_{DW}=1$. Since we wish to preserve gauge coupling unification, we will only consider additional $SU(3)$ states which are embedded in representations of $SU(5)$.  Because the PQ breaking scale is much smaller in these models, it will generate a $\mu$ term that is quite small.  In fact, one of the telling signatures of these types of models would be a TeV scale Higgsino. For the models of PQ breaking we will consider \cite{msy}, the $F$-terms for the PQ breaking are relatively large which leads to independent $\mu$ and $B$ terms.  However, since we will also couple these fields to additional representations of $SU(5)$, these models will also have modified gaugino mass spectra \cite{hiy,eo}.  In fact, the wino mass can be heavier than the bino mass.

Our paper is organized as follows: In the next section, we consider simple models within
the context of PGM with small $\mu$-terms. While we maintain sfermion mass universality,
we do allow for some non-universality in the soft Higgs masses in order to satisfy
both the EWSB conditions and the supergravity boundary condition on $B_0$. As we will see,
in the absence of a GM term, the above conditions are too restrictive to allow for solutions
with low $\mu$ (ie. a Higgsino LSP), even when the additional freedom afforded by non-universal
Higgs masses is allowed. Similarly, if the $\mu$ term is assumed to arise solely from a GM term,
the resulting relations between $B$ and $\mu$ prevents us from obtaining acceptable solutions. If instead
we allow for both a $\mu$ term in the superpotential and a GM term in the K\"ahler potential,
we can obtain interesting solutions with low $\mu$ and Higgsino dark matter. In this case, it is sufficient
to require a single non-universal Higgs mass ($m_1 = m_2$). In section 3, we will consider
models with a small $\mu$ term, and possible Higgsino dark matter, which arises from the dynamics of the theory. These models in particular incorporate a PQ symmetry and can also lead to an
axion and/or Higgsino dark matter candidate.
Our conclusions are given in section 4.

\section{PGM Models with low $\mu$}

One of the basic assumptions of PGM is that there are no fundamental singlets in the SUSY breaking sector.  This has little effect on the scalar masses, since they are generated in the K\"ahler potential and are proportional to operators composed of holomorphic and anti-holomorphic fields. The gaugino masses, on the other hand, come as corrections to the gauge kinetic function. The gauge kinetic function, which contains the gaugino fields, can only couple to operators forming singlets. Since there are no fundamental singlets in PGM, the gaugino masses are forbidden to leading order.  The leading order contribution is then a one-loop mass induced by anomalies.  Gaugino masses are thus much smaller than the sfermion masses.  The $A$-terms also tend to be quite suppressed without a singlet in the SUSY breaking sector and so can be ignored.

Here we will assume the sfermion masses are universal at the GUT scale, given by the gravitino mass, $m_{3/2}$. The GUT scale is defined by the renormalization scale at which the electroweak gauge couplings,
$g_1$ and $g_2$ are equal. The strong gauge coupling and the MSSM Yukawa couplings are all fixed at the weak scale. A full set of renormalization group equations are run between the weak and GUT scales.
We minimize the Higgs potential at the weak scale and the two minimization equations determine two of the parameters in the electroweak sector.  In the standard CMSSM with universal Higgs masses ($m_1 = m_2 = m_{3/2}$), the electroweak breaking conditions are used to determine $\mu$ and $B$  leaving $\tbt$ free.
In the more restrictive mSUGRA models \cite{bfs}, one must employ the boundary condition $B_0 = A_0 - m_{3/2}$ thus requiring a solution for $\tbt$ from the EWSB minimization conditions \cite{vcmssm}.
Somewhat more freedom is allowed if one allows one or both of the Higgs soft masses to
deviate from universality \cite{nonu}. These CMSSM-like models are known as NUHM1 (when $m_1 = m_2$)
\cite{nuhm1} or NUHM2 when both Higgs soft masses are allowed to be free \cite{nuhm2}.
In NUHM models, one can treat either $\mu$ or the Higgs pseudo-scalar mass $m_A$ as free parameters in the NUHM1,
or both $\mu$ and $m_A$ as free parameters in the NUHM2 and use the minimization conditions to solve for $m_1$ and/or
$m_2$.
In the analysis that follows, we will keep $\mu$ a free parameter and use the electroweak breaking conditions to determine two of the parameters $B_0$ (or equivalently a GM coupling),  $m_1^2$, and $m_2^2$.

In the simplest viable model of PGM, it is assumed that the Higgs bilinear mass $\mu$ had $R$-charge 2. In this case, the Higgs bilinear mass could be generated in the superpotential,
\begin{eqnarray}
W\supset c_H'm_{3/2} H_uH_d\equiv \mu_0H_uH_d\, ,
\end{eqnarray}
because $m_{3/2}$ has\footnote{ An explicit example of generating a $\mu$ proportional to $m_{3/2}$ is the axion model considered below. However, there are other possible ways.} $R$-charge 2. For this model where the $R$ charge of $H_uH_d$ is zero, a Giudice-Masiero term is also allowed
\begin{eqnarray}
K\supset c_HH_uH_d\, .
\end{eqnarray}
The Higgs bilinear terms are then
\begin{eqnarray}
&&\mu=c_Hm_{3/2}+\mu_0 \label{mu}\, , \\
&&B\mu=2c_Hm_{3/2}^2-\mu_0m_{3/2}\, .
 \label{bmu}
\end{eqnarray}
In this case, $B$ and $\mu$ are completely independent of each other and we have already imposed the
supergravity boundary condition $B_0 = -m_{3/2}$. Since we wish to keep $\mu$ a free parameter, we determine $B$ and Higgs soft masses using the EWSB conditions. For now, we will take $m_1=m_2$ (as in the NUHM1) at the GUT scale for simplicity. The conditions (\ref{mu}) and (\ref{bmu}) allow us to take the input value of $\mu$
(run up to the GUT scale) and the derived value of $B$ to determine $\mu_0$ and $c_H$.

In PGM models with full scalar mass universality, it is possible to obtain, a Higgs mass
compatible with the measured value for large $m_{3/2}$ (between $100$--$1500$ TeV) for a limited range in $\tbt = 1.7$--$3$ \cite{eioy,eo}. For fixed $\tbt$, as $m_{3/2}$ is increased, the Higgs mass is increased.
The increase in the Higgs  mass, $m_H$, becomes very rapid as $\mu$ decreases and we approach the focus point \cite{fp} region. However, with the exception of the extreme cases where $\mu \ll m_{3/2}$,
we are not able to find regions with acceptable Higgs masses and a Higgsino LSP simultaneously.

\begin{figure}[h]
\vskip 0.5in
\vspace*{-0.45in}
\begin{minipage}{8in}
\epsfig{file=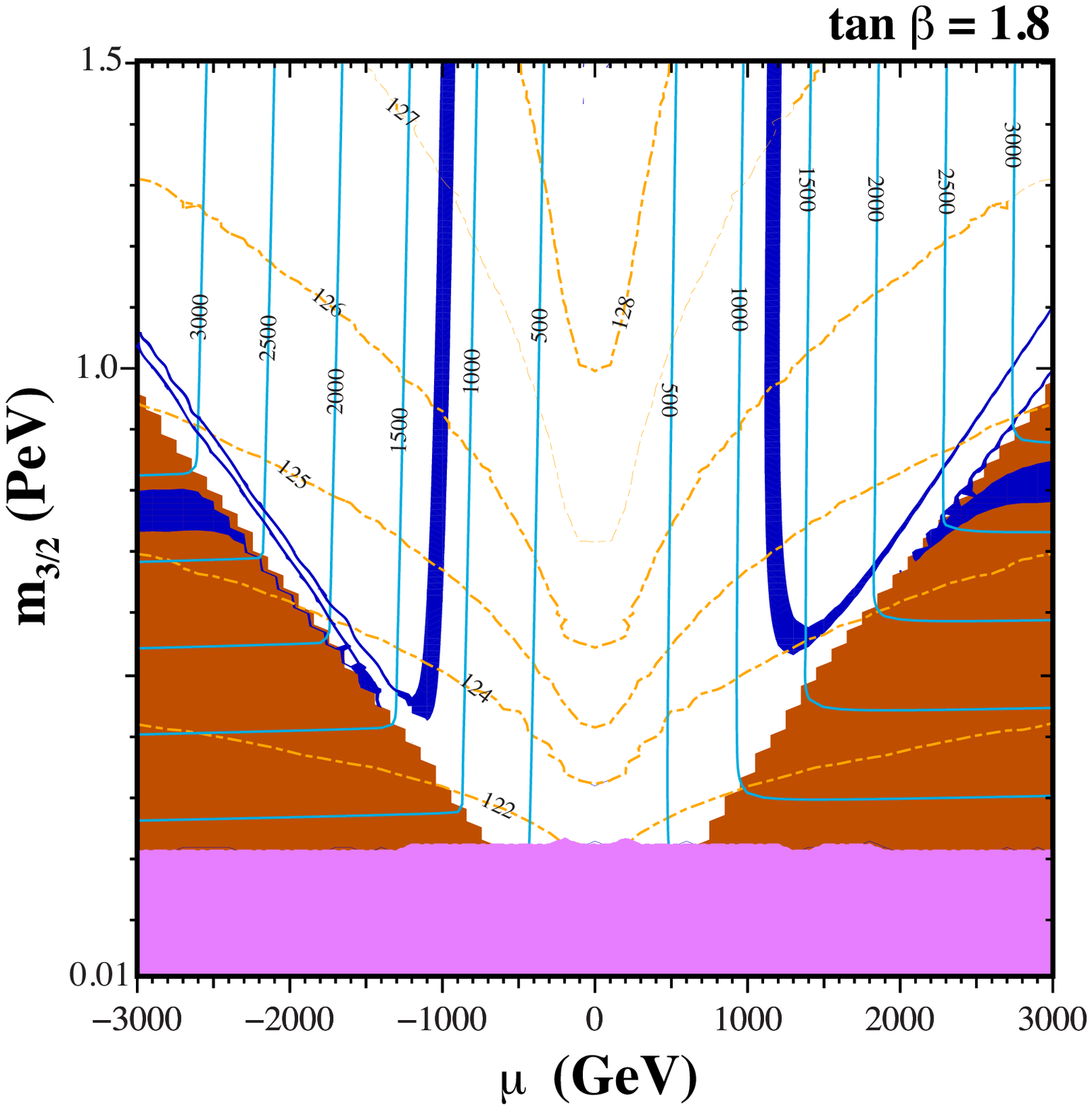,height=3.1in}
\epsfig{file=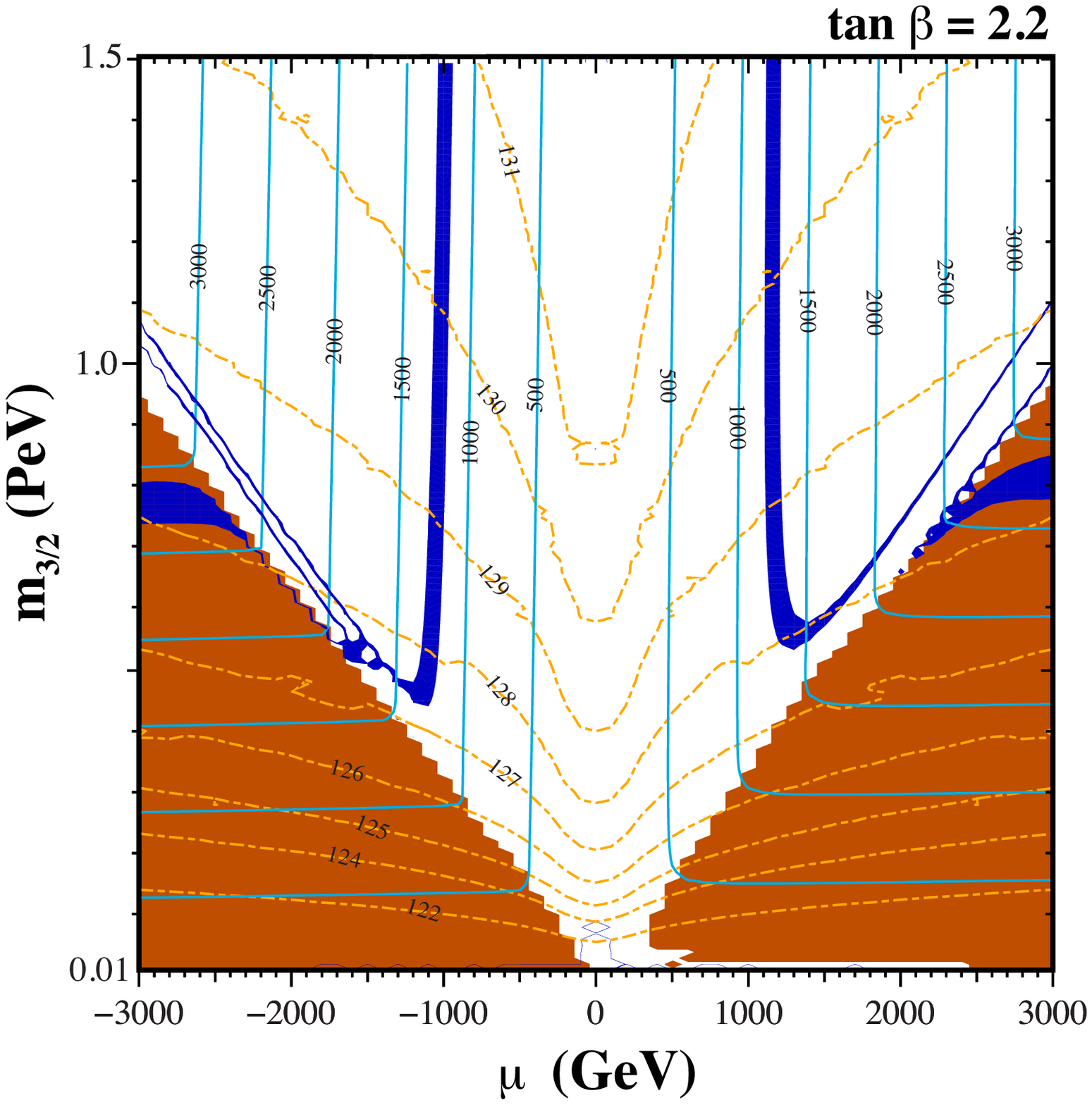,height=3.1in}
\hfill
\end{minipage}
\caption{
{\it
The $\mu, m_{3/2}$ plane for fixed $\tbt = 1.8$ (left) and 2.2 (right). The pink shaded region is excluded
as it contains a tachyonic stop. In the dark red shaded region, there is a wino LSP. In the remainder of the plane, the Higgsino is the LSP. Higgs mass contours, shown as orange dot-dashed curves, are given for
$m_H = 122, 124, 125, 126, 127, 128, 130$ and 132 GeV.  The light blue contours give the
LSP mass from 500-3000 GeV in 500 GeV intervals. When the LSP is a Higgsino, the contours are nearly vertical as the mass of the Higgsino depends primarily on $\mu$, whereas when the LSP is a wino, the contours are nearly horizontal as the anomaly mediated wino mass depends primarily on the gravitino mass.
}}
\label{fig:pgm1}
\end{figure}

Significantly more freedom is allowed in the NUHM even when the two Higgs soft masses are constrained to
be equal \cite{eioy2}.
In Fig.\,\ref{fig:pgm1}, we show examples of $\mu, m_{3/2}$ planes for fixed $\tbt = 1.8$ and 2.2 as labeled.
At each point on the plane, the EWSB conditions are used to solve for the $B$ term (or equivalently the GM coupling) and the Higgs soft masses (assumed here to be equal at the GUT scale, ie. NUHM1). In the left panel for $
\tbt = 1.8$, the pink shaded region with $m_{3/2} \la 200$ TeV, is excluded as the low energy spectrum contains a tachyonic stop.  In each panel, the dark red shaded regions, delineate the parts of the plane
with a wino LSP. In the complement, there is a Higgsino LSP. There are two sets of contours
in each panel. The orange dot-dashed contours correspond to a constant Higgs mass, and the
light blue contours show the values of the LSP mass across the plane. The GM coupling and the ratio of Higgs soft masses vary very
slowly with $m_{3/2}$ and are almost independent of $\mu$. For $m_{3/2} \simeq 200$ TeV, $c_H \approx
-0.5$ and $m_i/m_{3/2} \approx 1.5$, while for $m_{3/2} \simeq 1.5$ PeV, $c_H \approx -0.3$ and $m_i/m_{3/2} \approx 1.2$. For $\tbt = 2.2$, the GM couplings and Higgs masses also vary slowly.
For $m_{3/2} \simeq 100$ TeV, $c_H \approx
-0.2$ and $m_i/m_{3/2} \approx 1.2$, while for $m_{3/2} \simeq 1.5$ PeV, $c_H \approx -0.1$ and $m_i/m_{3/2} \approx 0.9$. Note that full scalar universality in this case occurs when $m_{3/2} \approx 530$ TeV
where the Higgs mass is somewhat larger than the observed mass
(for $\tbt = 1.8$, universality occurs at $m_{3/2} > 1.5$ PeV).

In the dark blue shaded regions,
the LSP has a relic density $\Omega h^2 = 0.11 - 0.13$ as preferred by recent Planck results \cite{Planck}.
The relic density in the Higgsino region depends only on the mass of the Higgsino \cite{osi} and the correct
relic density is obtained when $\mu \approx -1000$ GeV and 1200 GeV, with a Higgsino mass
just over 1200 GeV for both positive and negative $\mu$ (the difference between the Higgsino mass
and $\mu$ is due to one-loop threshold corrections).  These regions are seen as the
vertical strips in both panels.  At  large $|\mu|$, the wino is the LSP and
can have the correct relic density when its mass is approximately 2.7 TeV at $m_{3/2} \approx 800$ TeV.
In between these two extremes, the relic density can be obtained through coannihilation.
These coannihilation strips are seen as diagonals blue strips \footnote{Note that the relic density
in this region depends not only on the mass difference between the nearly degenerate
neutralino states, but also on their composition. The coannihilation strips are somewhat
offset from the gaugino/Higgsino degeneracy line because the next to lightest sparticle
becomes very mixed at slightly large $m_{3/2}$ (or smaller $|\mu|$) and leads to an enhanced annihilation cross section and hence the two coannihilation strips.}.

In this model, Higgsino dark matter with the desired relic density is possible for both positive and negative
values of $\mu$. For $\tbt = 1.8$, the region with acceptable dark matter occurs for values of the Higgs mass
$m_H \ga 123$ GeV and we find that $m_H = 126$ GeV when $m_{3/2} \simeq 1$ PeV. As $\tbt$ is increased, the Higgs mass increases rapidly. As one can see from the right panel of Fig.\,\ref{fig:pgm1},
when $\tbt = 2.2$, the Higgs mass is now greater than 127 GeV in regions with the correct relic density.
At larger $\tbt$, the Higgs mass is too large with respect to the experimental measurement.
Thus the model is restricted to a very narrow range in $\tbt$, but covers a large range in $m_{3/2}$
with values as low as $\sim 400$ TeV.

\begin{figure}[h!]
\begin{minipage}{8in}
\begin{center}
\hskip -1in
\epsfig{file=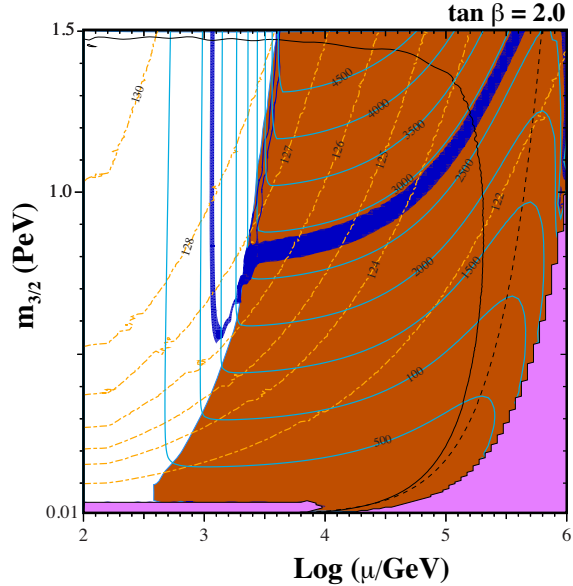,height=3.1in}
\hfill
\end{center}
\end{minipage}
\caption{
{\it
The $\log \mu, m_{3/2}$ plane for fixed $\tbt = 2.0$. The shadings and contour descriptions are as in
Fig.\,\ref{fig:pgm1}. Included here are also two thin black contours indicating the location
of scalar mass universality ($m_i/m_{3/2} = 1$) (solid) and $c_H = 0$ (dashed).
}}
\label{fig:pgm1.5}
\end{figure}

An extended view of the $\mu, m_{3/2}$ plane for $\mu > 0$ is shown in Fig.\,\ref{fig:pgm1.5} for $\tbt=2.0$.
As in the previous figure, the pink shaded region at very low $m_{3/2}$ is excluded due to a tachyonic
stop, and the pink shaded region in the lower right is excluded due to tachyonic pseudo-scalar.
Once again, we see a strip of Higgsino dark matter with $m_\chi \approx 1200$ GeV running
vertical near $\log \mu = 3.1$ This is connected to a thick strip of wino dark matter which runs towards
high $\mu$ and high $m_{3/2}$ with $m_\chi \la 3$ TeV. In between, we see again two strips near the border
between a Higgsino and wino LSP due to coannihilations.
Also plotted here is the contour where
we have full scalar mass universality.  This is seen as a black curve which is nearly horizontal near
$m_{3/2} = 1.5$ PeV and nearly vertical at $\log \mu \sim 5.2$. The Higgs soft masses are largest in the lower
left corner of the figure where they are approximately $1.3 m_{3/2}$.  To the right of the solid black contour,
the Higgs soft masses decrease rapidly and they are only about 20\% of the gravitino mass in the upper
right corner. The dashed black contour shows the position of the GM coupling $c_H = 0$.
It is largest at the far right where it is approximately $0.3$
and in the lower left where it is about $-0.3$.

Next, we look at another possible relationship between $\mu$ and $B\mu$.  If for example $H_uH_d$ has $R$-charge 2, the GM term is forbidden but the tree-level $\mu_0$ is not. This is the same as mSUGRA with $A_0=0$.  In this case, we get
\begin{eqnarray}
&&\mu=\mu_0\, , \\
&& B\mu=-m_{3/2}\mu_0\, .
\end{eqnarray}
In this scenario, $B$ cannot be varied independently of $m_{3/2}$ and there is no longer sufficient freedom in the EWSB conditions to be able to vary $\tbt$ and $\mu$ if full scalar mass universality is imposed. However, if we  allow both $m_1^2$ and $m_{2}^2$ to be free parameters, as in the NUHM2, the EWSB conditions can be met.  Therefore, we are able to maintain the supergravity boundary condition on $B_0$ and choose $\mu$ at the weak and set the Higgs soft masses at the weak scale so that the EWSB conditions are met.

\begin{figure}[h]
\vskip 0.5in
\vspace*{-0.45in}
\begin{minipage}{8in}
\epsfig{file=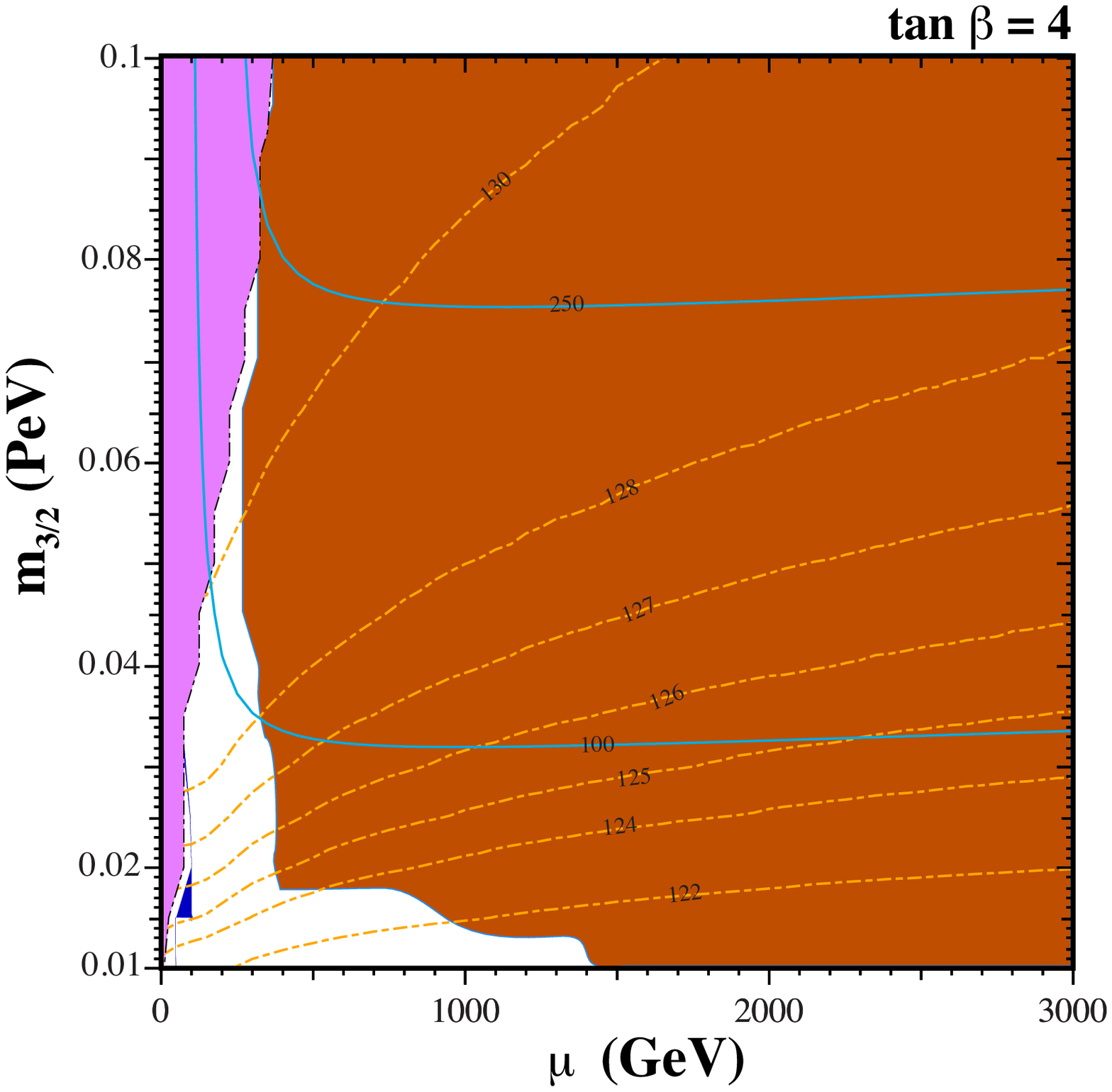,height=3.1in}
\epsfig{file=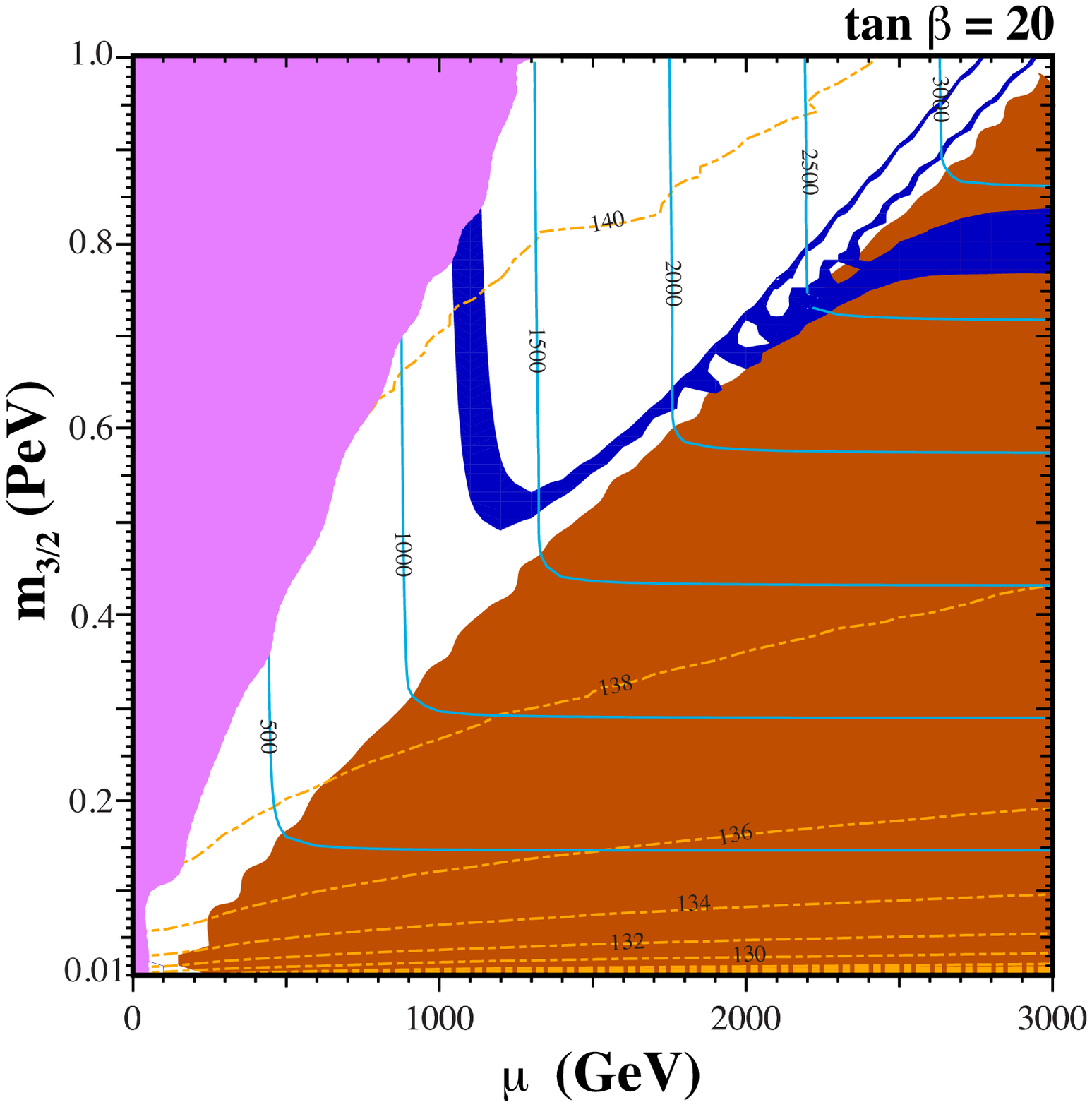,height=3.1in}
\hfill
\end{minipage}
\caption{
{\it
The $\mu, m_{3/2}$ plane for fixed $\tbt = 4$ (left) and 20 (right) with $c_H = 0$.
The pink shaded region is excluded
because the Higgs pseudoscalar is tachyonic. In the dark red shaded region, there is a wino LSP. In the remainder of the plane, the Higgsino is the LSP. Higgs mass contours shown as red dot-dashed curves as labeled.  The light blue contours give the
LSP mass as labeled.
}}
\label{fig:pgm2}
\end{figure}

Examples of the $\mu, m_{3/2}$ plane in this model are shown in Fig.\,\ref{fig:pgm2} for $\tbt = 4$ and $20$ as
labeled. In this case, the pink shaded region is excluded as it contains a tachyonic Higgs
pseudo-scalar. This region arises because the one-loop correction to Higgs potential is large driving $B$ to be very large when $\mu$ is small. To offset this large correction to $B$, the Higgs masses in the combination $m_1^2+m_2^2+2\mu^2$ needs to be taken quite negative.  Since this is the mass combination that sets the tree-level contribution to the pseudoscalar Higgs mass, $m_A$ tends to be tachyonic unless its one-loop corrections are large and positive. The other shadings and contours are the same as in Fig.\,\ref{fig:pgm1}. When
$\tbt = 4$, only a small portion of the parameter plane with relatively small $\mu$ and small $m_{3/2}$
admits a Higgsino LSP.  While we are able to obtain an acceptable Higgs mass in this region (as well as in
the wino LSP region), the relic density is too small to account for the observed density of dark matter (the
small blue shaded region at $\mu \sim 100$ GeV has a $\sim 10$ GeV mixed neutralino and chargino and is
not viable).
Note that this is not a fatal flaw in the model, but it does require additional fields
beyond the standard MSSM physics.  At larger $\tbt$ the LSP mass is increased and
we can find regions with the Planck inferred relic density. In the right panel of Fig.\,\ref{fig:pgm2},
we show the plane for $\tbt = 20$. Indeed we see regions with both Higgs and wino dark matter.
However, in this case, the Higgs mass is far too large and it lies between $138$ and $140$ GeV,
making this possibility unacceptable.

\begin{figure}[h]
\vskip 0.5in
\vspace*{-0.45in}
\begin{minipage}{8in}
\epsfig{file=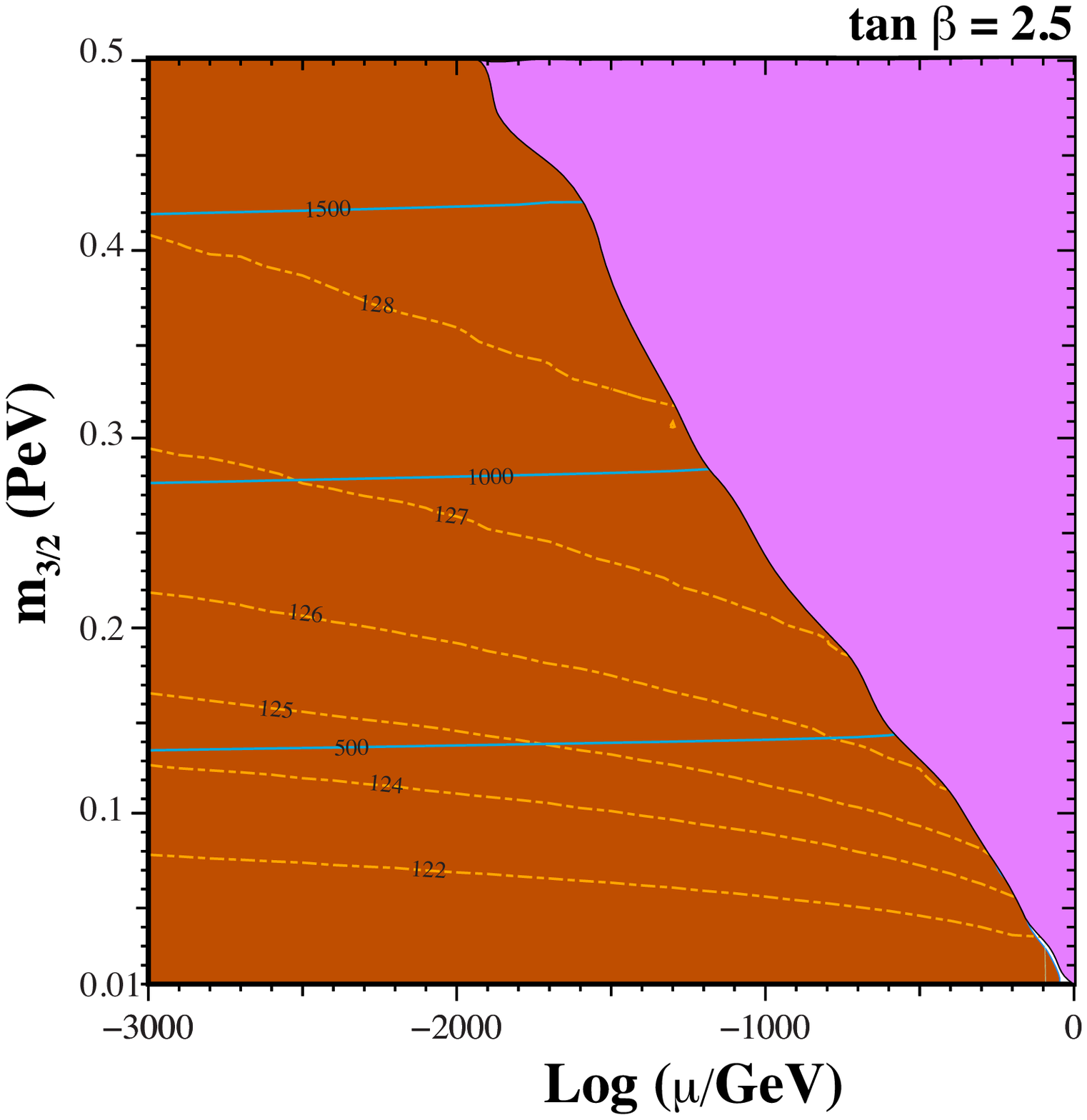,height=3.1in}
\epsfig{file=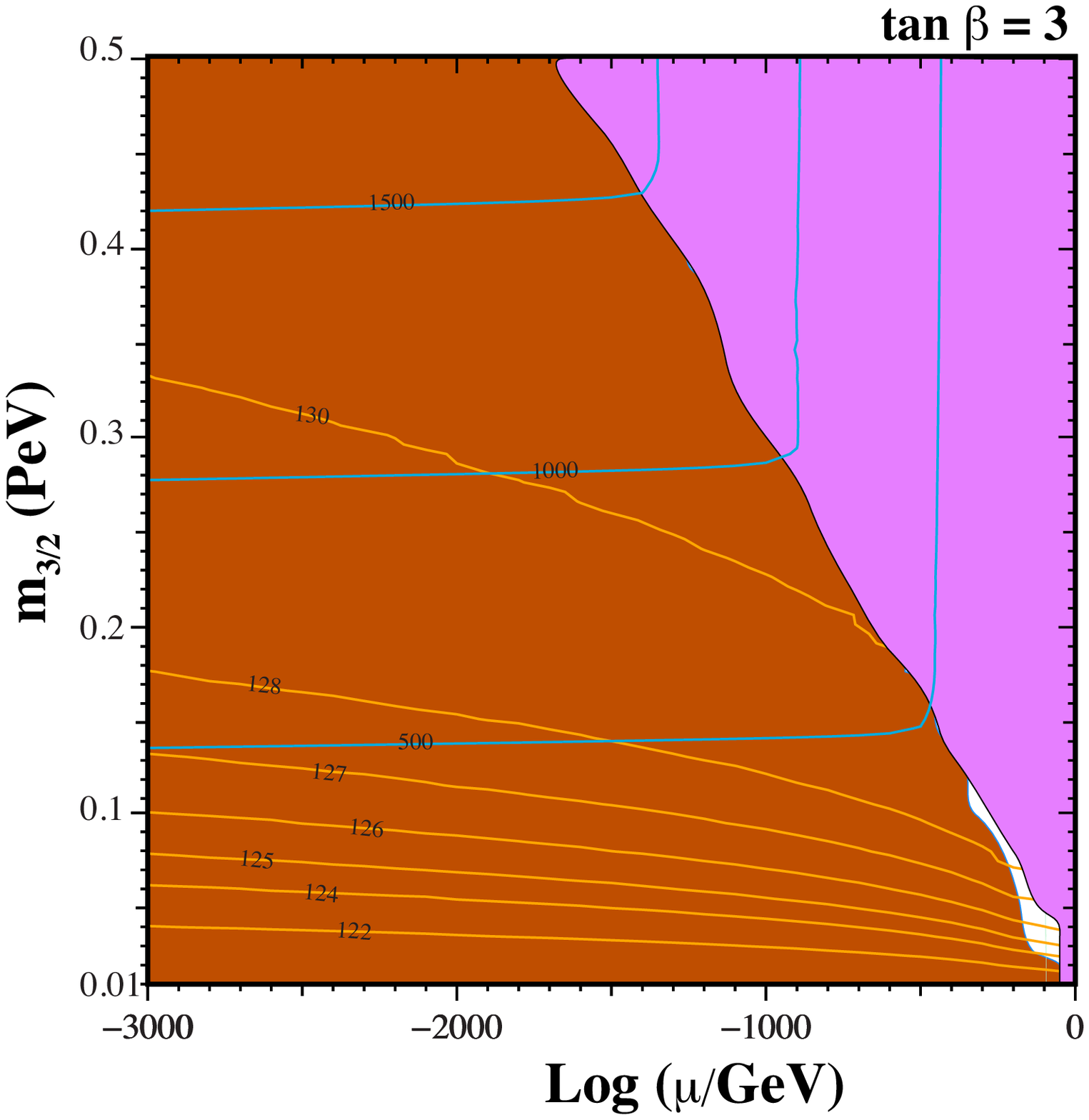,height=3.1in}
\hfill
\end{minipage}
\caption{
{\it
The $\mu, m_{3/2}$ plane for fixed $\tbt = 2.5$ (left) and 3 (right) with $\mu_0 = 0$.
The pink shaded region is excluded
because the Higgs pseudoscalar is tachyonic. In the dark red shaded region, there is a wino LSP. In the remainder of the plane, the Higgsino is the LSP. Higgs mass contours shown as red dot-dashed curves as labeled.  The light blue contours give the
LSP mass as labeled.
}}
\label{fig:pgm3}
\end{figure}

Another set of possible relationships for $B$ and $\mu$  comes from assuming $H_uH_d$ has $R$-charge zero, but exclude or suppress the coupling of $m_{3/2}$ to $H_uH_d$.  In this case the relationships for the EW parameters become
\begin{eqnarray}
&& \mu=c_H m_{3/2}\, ,\\
&& B\mu=2m_{3/2}\mu \, .
\end{eqnarray}
In this scenario, $B$ is also set by $m_{3/2}$.  As in the previous case, we will set $B$ and $\mu$ at the GUT scale and use the EWSB conditions to set the Higgs soft masses at the weak scale. This scenario is similar to the previous scenario. Since $\mu$ is very small and $B$ is quite large, the pseudoscalar Higgs tends to have a tachyonic mass in regions where the Higgsino could be the LSP and a good candidate for dark matter. These features can be seen in Fig.\,\ref{fig:pgm3}.

\section{Peccei-Quinn Breaking and $B$ and $\mu$}
As we have shown above, the models with a $\mu$-term around the TeV scale are viable even for
scalar masses in the hundreds of TeV to the PeV scale.
A TeV scale $\mu$-term like this can be achieved by, for example, assuming an approximate symmetry
which forbids the Higgs bi-linear term at leading order.
In general, this procedure eliminates a possible correlation between the
$\mu$-term and the gravitino mass $m_{3/2}$ and one of the salient features of pure gravity mediation,  that
there is only one additional scale other than the weak scale, is lost.  As we will see below, it is possible to rectify this short coming in theories with naturally small Higgs bi-linear terms.

One possible way to suppress the $\mu$ term is to assume that it
originates from the breaking of the Peccei-Quinn (PQ) symmetry \cite{Peccei:1977hh}
via a dimension five operator \cite{Kim:1983dt}.
Then, the size of the $\mu$-term is determined by the PQ-breaking scale $f_{PQ}$, i.e.
\begin{eqnarray}
\mu \simeq \frac{f_{PQ}^2}{M_{P}}\, .
\end{eqnarray}
To achieve a $\mu$-term at the TeV scale, we find that
\begin{eqnarray}
\label{eq:PQscale}
 f_{PQ} \simeq 5\times 10^{10}\,\left(\frac{\mu}{\rm TeV}\right)^{1/2}{\rm GeV}\, .
\end{eqnarray}

Interestingly, in some mechanisms for breaking the PQ symmetry, this range for the PQ-breaking scale can be related to the gravitino mass.
To illustrate this, let us start with the model of PQ-breaking in Ref.\,\cite{Murayama:1992dj}.
In this model, the PQ-symmetry is broken by the vacuum expectation values of the SM singlet fields $P$ and $Q$,
whose superpotential interactions include higher dimensional operators,
\begin{eqnarray}
W \ni \lambda\frac{P^3Q}{\Lambda} +g\frac{PQ}{\Lambda^\prime} H_uH_d \, . \label{WPQ}
\end{eqnarray}
Here,  $\lambda$ and $g$ are $O(1)$ coupling constants and
we have assumed that the PQ charge of $P$ is $-1$,  $Q$ is $3$, and $H_u$ and $H_d$ are $-1$.
In PGM models, the PQ-breaking fields obtain soft masses of $O(m_{3/2})$.
Then, if the squared soft masses of $P$ and $Q$
go negative (due to the renormalization group running effects of $O(1)$ Yukawa interactions)%
\footnote{The Yukawa couplings we refer to here are from either the right-handed neutrino coupling to $P$\,\cite{Murayama:1992dj}
or other matter fields coupling to $P$ or $Q$ (see Eq.\,(\ref{eq:additional})).}
the PQ-breaking scale is given by
\begin{eqnarray}
\langle P\rangle \sim
\langle Q\rangle \sim\sqrt{m_{3/2} M_{P}}  \sim 5\times 10^{11}\,{\rm GeV} \left(\frac{m_{3/2}}{100\,\rm TeV}\right)^{1/2}\, ,
\end{eqnarray}
if $\Lambda \sim \Lambda' \sim M_P$.
The resultant PQ-breaking scale is only one order of magnitude larger than the required scale found in Eq.\,(\ref{eq:PQscale}).
This separation can be easily removed if we further assume that the $P$ and $Q$ are composite fields
resulting from some strong dynamics having a dynamical scale at around the Planck scale.
In fact, by naive dimensional analysis\,\cite{Luty:1997fk,Cohen:1997rt}
the higher dimensional interactions between $P$ and $Q$ are enhanced,
while the origin of the $\mu$-term is intact, i.e. we expect that $(4\pi)^2 \Lambda \sim \Lambda^\prime \sim M_P$ and
\begin{eqnarray}
W\ni (4\pi)^2\lambda\frac{P^3Q}{M_{P}} +g\frac{PQ}{M_P} H_uH_d \, , \label{eq:WPQ2}
\end{eqnarray}
which leads to a suppressed PQ-breaking scale
\begin{eqnarray}
\langle P\rangle \sim \frac{1}{4\pi}\sqrt{m_{3/2} M_{P}}  \sim 5\times 10^{10}\,{\rm GeV} \left(\frac{m_{3/2}}{100\,\rm TeV}\right)^{1/2}\, .
\end{eqnarray}
In this way, the required PQ-breaking scale can be successfully interrelated to the gravitino mass.
It should be noted that the predicted $B$-term is $O(m_{3/2})$ with its exact size depending on the soft masses
of $P$ and $Q$ (see the appendix).

Let us note here that the axion predicted in this PQ-breaking model
contributes to the dark matter density.
In particular, axionic dark matter with the PQ-scale found in Eq.\,(\ref{eq:PQscale})
can account for all the dark matter if the domain wall number of the PQ-symmetry ($N_{DW}$) is
one and the PQ-breaking occurs after the end of inflation.
In this case, the axion density mainly comes from the decays of cosmic strings\,\cite{Hiramatsu:2012gg}
(see also Refs.\cite{Harari:1987ht,Chang:1998bq,Davis:1985pt,Battye:1993jv}).
This axion dark matter scenario is well motivated if the Hubble scale during inflation
is very high,  $H_I \sim 10^{13}$\,GeV, as it is in chaotic inflation models.
This scenario evades the severe constraints from the isocurvature fluctuation
in the CMB \cite{Visinelli:2014twa} and can be made consistent with isocurvature perturbations.%
\footnote{In the PQ-breaking model shown in Eq.\,(\ref{eq:WPQ2}), the isocurvature fluctuation can also be
suppressed if the vacuum expectation values of $P$ and $Q$ are very large during inflation
due to a negative Hubble mass squared\,\cite{Choi:2014uaa,Moroi:2014mqa}.
In these scenarios, a model with $N_{\rm DW} \neq 1$ is also viable since PQ-breaking occurs before
the end of inflation and the axion contribution to dark matter is only a few percent if the PQ-breaking
scale is as in Eq.\,(\ref{eq:PQscale}).
}
An ongoing joint analysis of the Planck and BICEP2 data sets may support
a high inflation scale if a non-negligible fraction of the detected $B$-mode polarization of BICEP2\,\cite{Ade:2014xna} is due to primordial gravitational waves.

It should be noted that it is easy to make the PQ-breaking model in Eq.(\ref{eq:WPQ2}) to have $N_{\rm DW} = 1$,
whereas the MSSM contributions is given by $N_{\rm DW} = 6$.
For example, let us introduce additional  ${\bf 5}$ and $\bar {\bf 5}$ pairs which couple to $P$ and $Q$,
\begin{eqnarray}
\label{eq:additional}
\Delta W=\lambda_P P {\bf 5}_3 \bar {\bf 5}_3 +\lambda_Q Q {\bf 5}_1 \bar {\bf 5}_1+\lambda_Q Q {\bf 5}_2 \bar {\bf 5}_2\, ,
\end{eqnarray}
where we have taken the couplings to $Q$ to be universal for simplicity.
In this case, the contributions of the additional matter to the domain wall number is ${\mit\Delta }N_{\rm DW} = -5$,
which leads to $N_{\rm DW} = 1$.

Finally, let us discuss the effects of the additional ${\bf 5}$ and $\bar {\bf 5}$ on the gaugino masses. As the theory is run past the scale $\lambda_P P$, for example, the ${\bf 5}$ and $\bar {\bf 5}$ are decoupled and likewise for the states coupled to $Q$. After all the new SM charged fields have decoupled, the coefficients of beta functions for the gauginos return to their SM value. However, there is also an additional threshold correction from integrating out these states which is added to the gaugino masses at the scale $\lambda_{Q,P}(Q,P)$. This contribution is present because the $F$-terms of $Q$ and $P$ are non-zero.  Although we added the full threshold correction given in \cite{bpmz}, an easier to understand and very accurate approximation can be found. Because the $F$-terms of $P$ and $Q$ split the scalar masses of the ${\bf 5}$ and $\bar {\bf 5}$\footnote{In the approximate equalities below we have neglected $m_{P}$ and $m_Q$ because $\langle P \rangle , \langle Q \rangle  \gg m_{3/2}$.},
\begin{eqnarray}
m_{5_{3\pm}}^2=\lambda_P^2\langle P \rangle ^2+m_P^2\pm \lambda F_P\simeq \lambda_P^2\langle P \rangle ^2\pm \lambda_P F_P\, ,\\
m_{5_{1,2\pm}}^2=\lambda_Q^2 \langle Q \rangle ^2+m_Q^2\pm \lambda F_Q\simeq \lambda_Q^2\langle Q \rangle ^2\pm \lambda_Q F_Q\, ,
\end{eqnarray}
there is a one-loop correction to the gaugino masses with the additional $\five$ and $\fivebar$'s acting as messengers.  Using the standard gauge mediation calculation, we find that this contributes
\begin{eqnarray}
\Delta M_i= \frac{g_i^2}{16\pi^2}\left( \frac{F_P}{\langle P \rangle }+2\frac{F_Q}{\langle Q \rangle }\right)\, ,
\end{eqnarray}
to each of the  gaugino masses.

As should be clear from the discussion above, PQ symmetry breaking requires the mass squared of $P$ to run
negative.  This can be achieved either through the coupling $\lambda_P$ in Eq. (\ref{eq:additional}) or
as in the original model developed in \cite{msy} which couples $P$ to right-handed neutrinos through
\beq
W \ni h_N P N N\, .
\eeq
In this case, not only does the coupling $h_N$ help drive $m_P^2 < 0$, thus generating a vacuum expectation value for $P$, but
this term also supplies a Majorana mass for $N$ of the right size enabling the see-saw mechanism for neutrino masses.
A similar model was recently considered in the context of the little hierarchy problem \cite{howie}.
In the results presented below, we will use $h_N$ as the dominant mechanism for driving PQ breaking,
but the coupling $\lambda_P$ would play essentially the same role.
If in addition, the coupling $\lambda_Q$ in Eq. (\ref{eq:additional}) were large, the RG evolution of
$m_Q^2$ would also run negative.  However, in this case, there is the danger of generating unbounded from below directions if these masses are too negative.  In fact if $m_{Q}^2<0$, the direction $|Q|=|H_u|=|H_d|=0$ is unbounded from below with no metastable point in the potential. The other possibly problematic direction is $|Q|=|H_2^+|=\lambda P^2+gH_u^0H_d^0=0$, $|H_u^0|=\pm|H_d^0|$ and $P\to \infty$. This direction is problematic at tree-level if
\begin{eqnarray}
m_{H_u}^2+m_{H_d}^2\pm \frac{g}{\lambda}m_P^2<0 \,.
\end{eqnarray}
However, if the one-loop Coleman-Weinberg potential for the Higgs sector is added, the potential becomes metastable as long as the vacuum expectation value of the Higgs  can be stabilized near the weak scale. For this reason,
we choose to take $\lambda_Q$ small, and set its value equal to $10^{-3}$ thus ensuring that $m_Q^2>0$.
Note that $Q$ picks up a vacuum expectation value due to an effective linear term as explained in the appendix.

\begin{figure}[ht!]
\vskip 0.5in
\vspace*{-0.45in}
\begin{minipage}{8in}
\epsfig{file=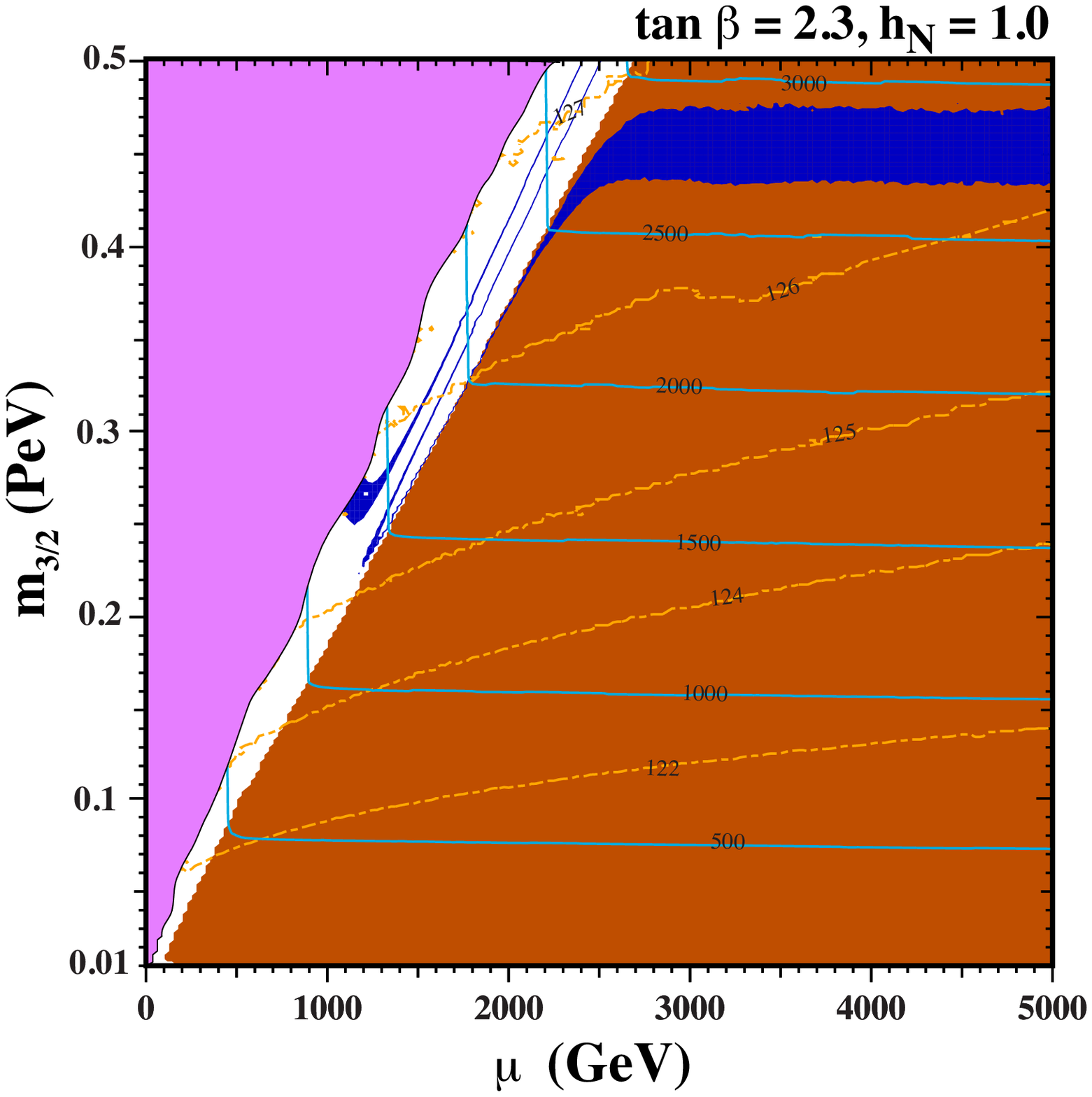,height=3.1in}
\epsfig{file=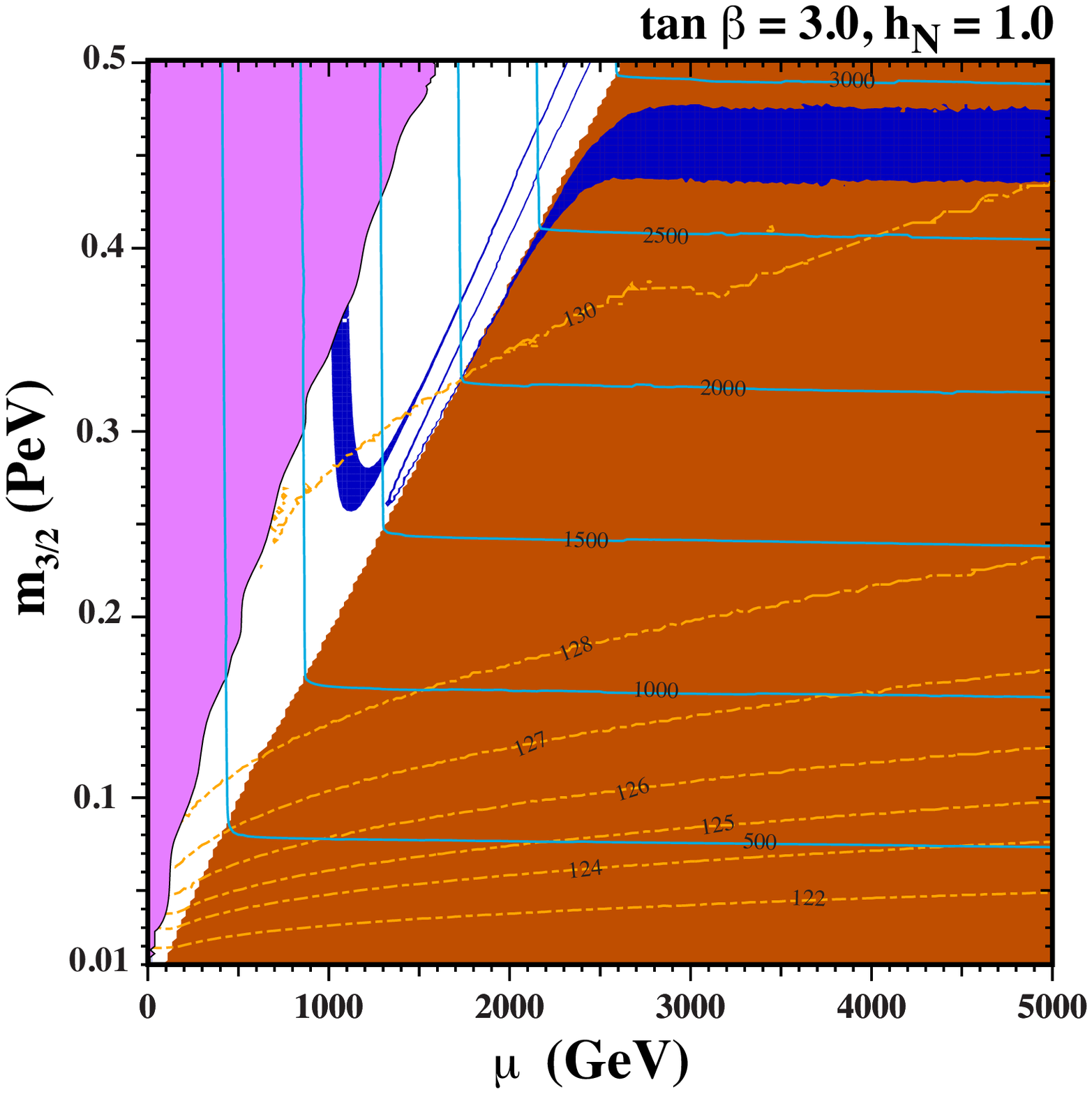,height=3.1in}
\hfill
\end{minipage}
\caption{
{\it The $\mu, m_{3/2}$ plane for fixed $h_N = 1$ and  $\tbt = 2.3$ (left) and 3.0 (right). The pink shaded region is excluded
because the Higgs pseudoscalar is tachyonic. In the dark red shaded region, there is a wino LSP.
Higgs mass contours shown as red dot-dashed curves are given for
$m_H = 122, 124, 125, 126, 127, 128$ and $130$ GeV.  The light blue contours give the
LSP mass from 500-3000 GeV in 500 GeV intervals.}}
\label{fig:VarTb}
\end{figure}

Another complication of these types of models is EWSB.  As discussed earlier, the PQ theory generates EWSB parameters with size
\begin{eqnarray}
\mu\sim \frac{m_{3/2}}{16\pi^2}\,, \quad \quad \quad B\sim m_{3/2}\, .
\end{eqnarray}
Because we are considering PGM, $m_{3/2}\gg M_Z$, in order to get an electroweak scale vacuum expectation value
for the Higgs, we need the magnitude of the determinant of the Higgs mass matrix to be at most of order $m_{3/2}^2M_Z^2$.  This amounts to $m_1^2m_2^2-(B\mu)^2 \ll m_{3/2}^4$. There is only one way to accomplish this with $B\sim m_{3/2}$ and $\mu\ll m_{3/2}$, the product of the Higgs masses need to be smaller than $m_{3/2}$.  This tends to be problematic since this gives a small or negative tree-level mass for the pseudoscalar, $m_A^2=m_1^2+m_2^2+2|\mu|^2$, and the one-loop corrections may not help. This effectively places a lower bound $\mu$ for a given value of $m_{3/2}$. This behavior can be seen in Fig.\,\ref{fig:VarTb} where larger values of $m_{3/2}$ require larger values of $\mu$ to avoid a tachyonic Higgs
pseudoscalar.

In Fig.\,\ref{fig:VarTb}, we show two examples of the $\mu, m_{3/2}$ plane.
Here, we have chosen $\lambda_P = 1$, $\lambda_Q = 10^{-3}$.  As explained in the appendix,
the axion decay constant, is proportional to $(m_{3/2}/\lambda)^{1/2}$, where $\lambda$ is the non-renormalizable coupling in Eq. (\ref{WPQ}). In the results that follow, we have set the axion
decay constant to be $F_{PQ} = 5 \times 10^{10}$ GeV corresponding to an axion relic density
sufficient to account for the dark matter.  For $m_{3/2} = 100$ TeV, we require $\lambda = 0.19$.
Thus we should in principle require the LSP relic density
to be small. However, our results for the supersymmetric masses are only weakly sensitive to the choice of the axion decay constant, and lowering the constant (by increasing $\lambda$) will not change the qualitative picture. In the left panel of Fig.\,\ref{fig:VarTb}, we take $\tan \beta = 2.3$ and we take $\tan \beta = 3.0$ in the right panel. In both cases, we assume $h_N = 1.0$.

The shadings and contours in Fig.\,\ref{fig:VarTb} are the same as in earlier figures.
In the red shaded region, the wino is the LSP and the neutralino mass contours are horizontal.  In the
unshaded area, the Higgsino is the LSP and the neutralino mass contours are vertical.
As before, the blue shaded regions show the area where the neutralino relic density lies between
$\Omega_\chi h^2 = 0.11$--$0.13$. If axions make up the dark matter, then viable regions in the parameter
space lie below the blue shaded area where the relic density is below the cosmic density.
However, as noted above,  if we lower the axion decay constant, then the blue shaded regions
correspond to neutralino dark matter. Thus in this model, we can have either
axion dark matter, neutralino dark matter, or a mixture of the two.%
\footnote{See \cite{axhiggs} for models with
similar possibilities.}
 For $\tan \beta = 2.3$, there is only a small region where the
dark matter can be a Higgsino (at higher values of $m_{3/2}$ there is a tachyonic Higgs pseudo-scalar).
For larger $\tan \beta$, more of the Higgsino strip is visible,  though at an excessive value of the Higgs mass.
In both panels, we again see a region where coannihilation determines the relic density. Between
these two narrow (diagonal) strips, the relic density is small.

As one can see from comparing the two panels in Fig.\,\ref{fig:VarTb}, the Higgs masses increase with increasing $\tan\beta$.  Unless $\tan\beta$ is relatively small, the Higgs mass will be larger than the $126$ GeV measured at the LHC experiment.  Second, the size of the tree-level contribution to the pseudoscalar mass, $m_A=\sqrt{2B\mu/\sin2\beta}$, increases as $\tan\beta$ increases.  As a result, the value of $\mu$ necessary to get a positive mass squared for the pseudoscalar is smaller.  This behavior is reflected in the figures where we see that the area of the pink shaded (excluded) region is smaller at higher $\tan \beta$. If $\tan\beta < 2.3$, the regions with Higgsino dark matter are ruled out.

\begin{figure}[t!]
\vskip 0.5in
\vspace*{-0.45in}
\begin{minipage}{8in}
\epsfig{file=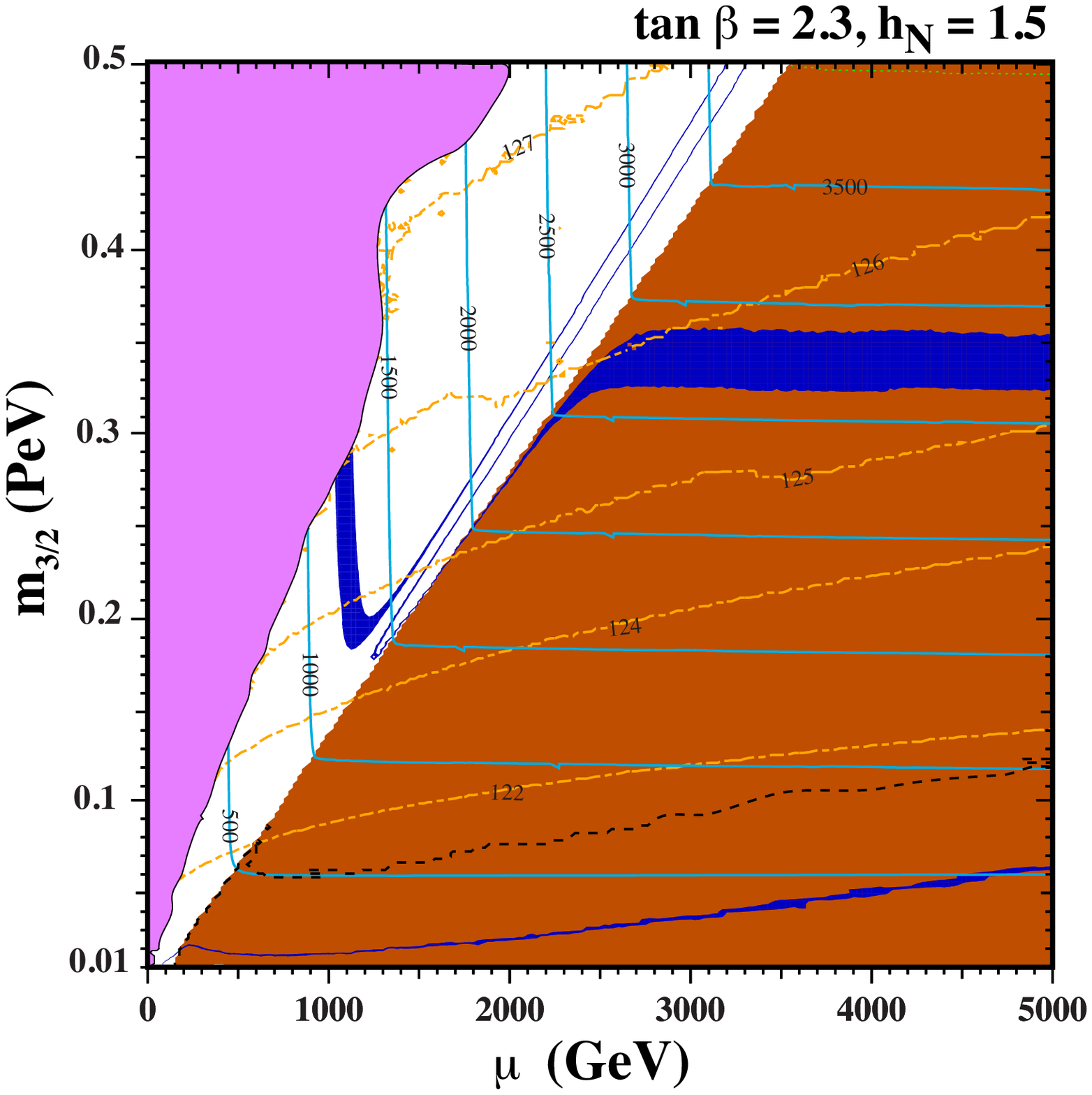,height=3.1in}
\hskip -0.in
\epsfig{file=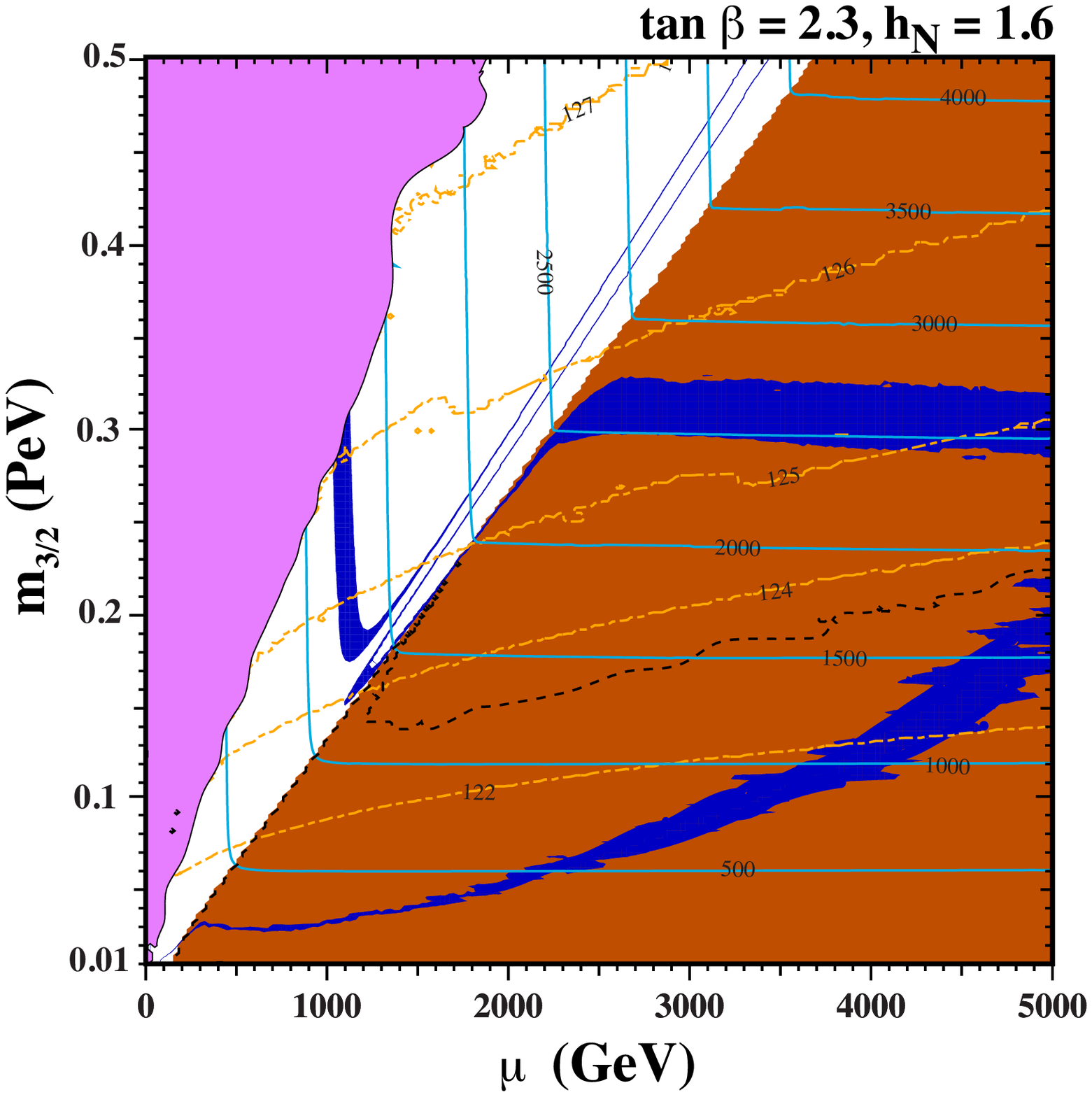,height=3.1in}\\
\epsfig{file=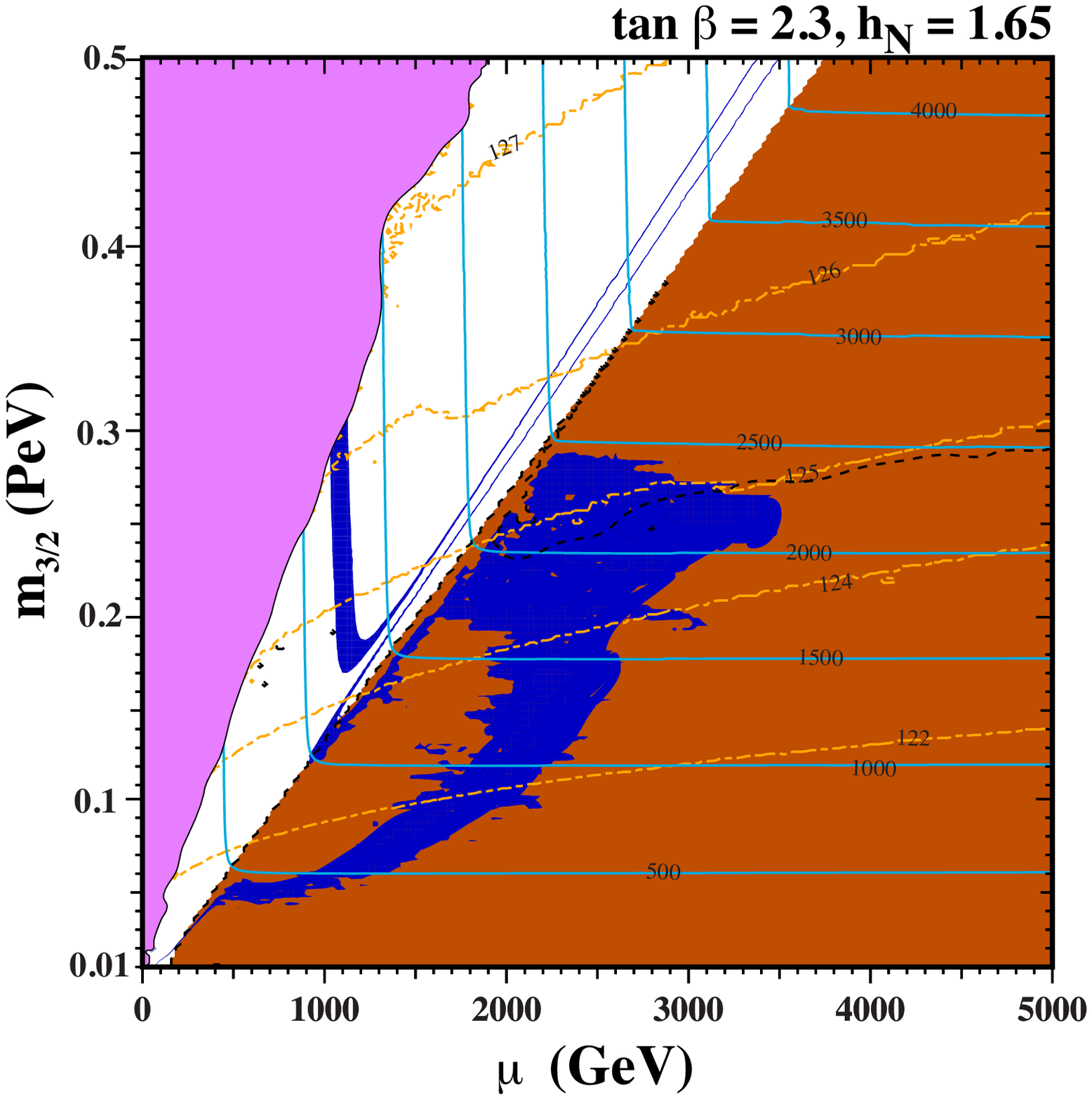,height=3.1in}
\epsfig{file=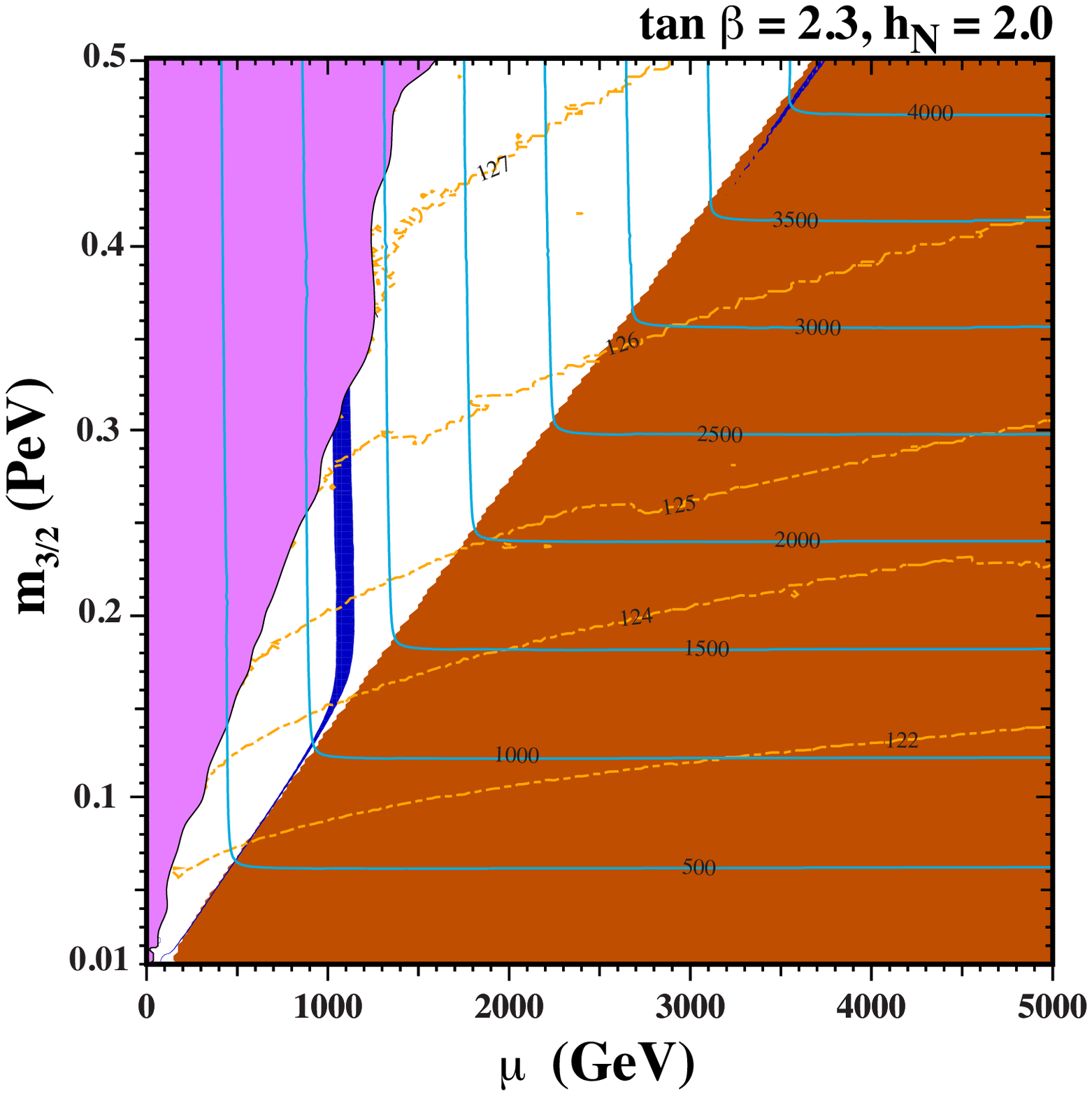,height=3.1in}
\hfill
\end{minipage}
\caption{
{\it
 The $\mu, m_{3/2}$ plane for fixed $\tbt = 2.3$ and $h_N=1.5$ (top left), $h_N=1.6$(top right), $h_N=1.65$ (bottom left) and $h_N=2$ (bottom right). The pink shaded region is excluded
because the Higgs pseudoscalar is tachyonic. In the dark red shaded region, there is a wino (bino) LSP for regions above (below) the black dashed line. Higgs mass contours shown as red dot-dashed curves are given for
$m_H = 122, 124, 125, 126$ and $127$ GeV.  The light blue contours give the
LSP mass from 500-4000 GeV in 500 GeV intervals. }}
\label{fig:VarHN}
\end{figure}

As noted above, the coupling of the $P$ field to right handed neutrinos
plays an important role in determining the soft mass parameter of $P$ and ultimately on the gaugino
masses. In Fig.\,\ref{fig:VarHN}, we vary $h_N$ between 1.5 and 2.0 as labeled.
As it turns out, variations in $\lambda_P$ give very similar results. Variation in $\lambda_Q$ tend to produce little variation unless $\lambda_Q$ is order one, in which case, it tends to drive $m_Q^2<0$ which has no stable minimum.  So we will only present figures for variation in $h_N$.

 For small $h_N$, the $\mu, m_{3/2}$ plane  is qualitatively similar to the case where $B=-m_{3/2}$ (seen in Fig.\,\ref{fig:pgm2}). For small $h_N$, $m_P^2$ is not driven very negative and so the vacuum expectation value of $P$, for  a given value of $\lambda$ tends to be small.  This results in smaller threshold corrections from integrating out the additional $\five$ and $\fivebar$ pairs and so we are in a case similar to that for $B=-m_{3/2}$.  As $h_N$ increases, $m_P^2$ becomes more negative and so the vacuum expectation value of $P$ becomes larger, for a given value of $\lambda$.  This greatly increases the size of the threshold corrections from the additional $\five$ and $\fivebar$ pairs.  These corrections become so large that they completely cancel the anomaly mediated contribution to the wino mass.  If $h_N$ is further increased, the wino mass becomes negative and can be larger than the bino and/or Higgsino.  Because the two Higgsinos, the wino, and the bino are nearly degenerate for this case, we get a rich dark matter behavior we see in
 Fig.\,\ref{fig:VarHN}.

 In all four panels of  Fig.\,\ref{fig:VarHN}, we see that
the Higgsino dark matter
 region extends down below the excluded pink shaded region and that points in that region have
 acceptable Higgs masses between 124 and 126 GeV. In all panels, we have a viable Higgsino
 dark matter candidate. Below that region, the Higgsino density is too small and we have axion dark matter.%
\footnote{As noted earlier, in order to have Higgsino dark matter, we would be required to lower
 the axion decay constant. When this is lowered by a factor of 10, we would require $h_N \simeq 1.3$
 to reproduce the panel displayed with $h_N = 1.6$. Higher and lower values of $h_N$ can be found
 to reproduce the other panels with a smaller axion decay constant. Similarly the panel with
 $h_N = 1.6$--$1.65$ can be approximately reproduced with $h_N = 1$ and $\lambda_P = 1.8$--$1.85$.}
 When $h_N$ = 1.5,
 we see in the gaugino LSP region (shaded red), a new strip where the relic density is achieved
 with a bino LSP. Below this strip (at $m_{3/2} \sim 50$ TeV, the relic density is too high.
 As $m_{3/2}$ is increased, the wino and bino masses become more degenerate and the
 relic density drops.  Above the dashed curve, the wino becomes the LSP.  As $h_N$ is increased,
 we see that the bino and wino relic density regions approach and merge when $h_N = 1.65$
 and have effectively disappeared when $h_N = 2.0$ where the gaugino region is now entirely bino-like.

\section{Summary}
Pure gravity mediation models are among the simplest viable phenomenological models of
supersymmetry on the market.  While much of the spectrum is quite heavy (the scalars
are at the multi TeV to PeV scale), gaugino masses may still lie within reach of the upcoming
run at the LHC.  These models incorporate radiative electroweak symmetry breaking, and can
accommodate a Higgs mass in the experimentally measured range of $124$--$126$ GeV.
In their most simple form, PGM models predict that the wino is the LSP and may
yield the correct relic density when the gravitino mass is of order half a PeV.
These results require a relatively narrow range for $\tan \beta$ around 2.

The minimal viable PGM model with two free parameters can easily be extended by allowing
non-universal Higgs masses. Here, we have exploited this possibility to examine
models with a relatively light (order 1 TeV) Higgsino.
By allowing the two Higgs soft masses ($m_1 = m_2$) to differ from the gravitino mass
at the GUT scale, we are able to allow $\mu$ to be a free parameter. For $\mu \ll m_{3/2}$,
the Higgsino may become the LSP and a viable dark matter candidate.
In this case, the charged Higgsino is nearly degenerate with the LSP, and
the discovery of a $\sim 1$ TeV chargino along with a missing energy signal
and no other superpartners, may point to models of this type.

The $\mu$ parameter is a known enigma in supersymmetric models, and
here we have also extended the PGM model to incorporate Peccei-Quinn
symmetry breaking which can explain the origin of the $\mu$ term.
By coupling the PQ fields to a right handed neutrino sector, the origin
of the large Majorana mass needed for the neutrino see-saw may also be explained.
This model is rich with dark matter possibilities as it may either
posses an axion dark matter candidate, a neutralino dark matter candidate
in the form of a Higgsino, wino, or bino, or some mixture of {\em all} of the above.

\section*{Acknowledgments}
The work of J.E. and K.A.O. was supported in part
by DOE grant DE-SC0011842 at the University of Minnesota.
The work  of M.I and T.T.Y  was supported by the Grant-in-Aid for Scientific research from the Ministry of Education, Science, Sports, and Culture (MEXT), Japan No. 26104009/26287039. and No. 24740151/25105011
and by the World Premier International Research Center Initiative (WPI), MEXT, Japan.
The work of M.I. was also supported by the Japan Society for the Promotion of Science (JSPS) No. 26287039 (M.I.).

\appendix
\section{Peccei Quinn Model}
Here we discuss the details of the PQ breaking sector. The PQ breaking portion of the superpotential is given by
\begin{eqnarray}
W=\frac{\lambda}{\Lambda}P^3Q+\frac{g}{M_P}H_uH_dPQ +h_N P N N\label{RHN}\, ,
\end{eqnarray}
where $\Lambda=M_P/(4\pi)^2$. The PQ charge of $P$ is $-1$ and for $Q$ it is $3$. We have neglected the Yukawa couplings for the fields $P$ and $Q$ with the additional $\five$ and $\fivebar$'s which can be found in the text. From here on, we will also neglect the Higgs fields as they will be a small perturbation to what we consider here. If we now include supersymmetry breaking affects, we get the PQ breaking potential
\begin{eqnarray}
V= 9|\lambda|^2\frac{|P^2Q|^2}{\Lambda^2} + |\lambda|^2\frac{|P|^6}{\Lambda^2} +m_P^2 |P|^2+m_Q^2 |Q|^2+\lambda\frac{m}{\Lambda}\left(P^3Q+h.c.\right)\, .
\end{eqnarray}
If the Yukawa couplings of $P$ are large enough, the RG running will drive $m_P^2$ negative inducing $PQ$ breaking. Once a vacuum expectation value of $P$ is generated, the term proportional to $m$ is effectively a linear term for $Q$. This effective linear term for $Q$ destabilizes the origin for $Q$, and as we will see below, gives $Q\sim P$.  From here on out we will assume $P$ and $Q$ are real. The minimum is found by solving the equations
\begin{eqnarray}
&&V_P=36\lambda^2\frac{P^3Q^2}{\Lambda^2}+6\lambda^2\frac{P^5}{\Lambda^2} + 2m_P^2P+6\lambda\frac{m}{\Lambda}P^2Q=0\,,\\
&& V_Q=18\lambda^2\frac{P^4Q}{\Lambda^2}+2m_Q^2Q+2\lambda\frac{m}{\Lambda}P^3=0\,.
\end{eqnarray}
To simplify the expressions we take $m_P^2=-xm_{3/2}^2$ and $m_Q^2=ym_{3/2}^2$. We also take $m=m_{3/2}$ which is set by mSUGRA. We now solve for the vacuum expectation values at the minimum.  There, solutions take the form
\begin{eqnarray}
P=Z\sqrt{\frac{m_{3/2}\Lambda}{\lambda}}\,,
\quad \quad Q= -\frac{Z^3}{9Z^4+y}\sqrt{\frac{m_{3/2}\Lambda}{\lambda}}\,, \label{PQE}
\end{eqnarray}
where $Z$ is the solution of
\begin{eqnarray}
243Z^{12}+(-81x+54y-9)Z^8+(-18xy+3y^2-3y)Z^4-xy^2=0\, .
\end{eqnarray}
Generically, $Z$ will be of order one and we expect the PQ breaking scale to be of order $F_{PQ}\sim \sqrt{m_{3/2}\Lambda}$ for $\lambda\sim 1$.

The expression for $\mu$ can be written as
\begin{eqnarray}
\mu=-\frac{m_{3/2}}{(4\pi)^2}\frac{g}{\lambda} \frac{Z^4}{3Z^4+1}\,.
\end{eqnarray}
From this we see that the order one variations of $\mu$ are set by $g$ and $\lambda$.  However, $\lambda$ is set by imposing constraints on $f_{PQ}$ and so $g$ determines $\mu$.   The PQ breaking scale is determined from the equation $f_{PQ}=\sqrt{P^2+9Q^2}$.  For example, for $Z=1/2$ we find
\begin{eqnarray}
f_{PQ}=5\times 10^{10} {\rm GeV}\left(\frac{m_{3/2}}{10^5 {\rm GeV}}\right)^{1/2}\left(\frac{0.19}{\lambda}\right)^{1/2}\,.
\end{eqnarray}
For $Z=1/2$, we also get
\begin{eqnarray}
\mu=-131 ~{\rm GeV} \left(\frac{g}{1}\right)\left(\frac{0.19}{\lambda}\right) \left(\frac{m_{3/2}}{10^5 ~{\rm GeV}}\right)\,.
\end{eqnarray}

Next we look at the $N_{DW}$. According to \cite{Geng:1990dv}, the instanton breaks the $U(1)_{PQ}$ to a $Z_N$ where
\begin{eqnarray}
N=\left|\sum\limits_{i=1}^{N_g} 2Q_i+u_i+d_i\right|\,.
\end{eqnarray}
In the case of $SO(10)$ unification this reduces to
\begin{eqnarray}
N=12|Q_{SM}|\,,
\end{eqnarray}
where $Q_5=Q_{10}=Q_{SM}$ is the PQ charge of the SM fields and we get $N=6$ for $Q_{SM}=\pm 1/2$.

To verify that we do indeed have a $Z_N$ symmetry we discuss the transformations of the different gauge invariant operators we consider. Each gauge invariant operator will transform as
\begin{eqnarray}
\Phi\to e^{i2\pi Q_{\Phi}/N} \Phi\,,
\end{eqnarray}
under the $Z_N$ symmetry where $\Phi$ is any gauge invariant operator charged under the PQ symmetry. To understand how $N$ relates to $N_{DW}$, we examine the gauge invariant operators: $H_uH_d$, $P$, and $Q$.  These operators have $PQ$ charges $Q_P=-1$, $Q_Q=3$, and $H_uH_d$ has charge 2, giving
\begin{eqnarray}
&&\langle P \rangle \to e^{i2\pi Q_P/N} \langle P \rangle = e^{-i2\pi/N} \langle P \rangle\,, \\
&&\langle Q \rangle \to e^{i2\pi Q_Q/N} \langle Q \rangle = e^{i6\pi/N} \langle Q \rangle\,, \\
&&\langle H_uH_d \rangle \to e^{i2\pi(Q_{H_u}+Q_{H_d})/N} \langle H_uH_d \rangle = e^{-i4\pi/N} \langle H_uH_d \rangle\, .
\end{eqnarray}
From these equations, we see that $Q$ transforms under a $Z_{N/3}$ subgroup of $Z_N$, and $H_uH_d$ transforms under a $Z_{N/2}$ subgroup of $Z_N$.   However, because $P$ transforms under the full $Z_N$ discrete symmetry, we have $N_{DW}=N$ and so for $Q_{16}=1/2$ we have $N_{DW}=6$.

$N_{DW}$ can be changed by coupling additional colored states to $P$ and $Q$. There are many possible choices if we allow for non-renormalizable operators.  However, if we restrict our choices to only marginal couplings we find
\begin{eqnarray}
 \lambda_Q(\bar 5_1 5_1 + \bar 5_2 5_2)Q +\lambda_P P \bar 5_3 5_3\nonumber\,,
\end{eqnarray}
which will give $N_{DW}=1$.

Now we examine $B$ terms in these models. The supergravity potential generates a soft breaking term of
\begin{eqnarray}
-{\cal L}_{soft}\supset m_{3/2}g\frac{PQ}{M_P} H_uH_d+h.c.=\mu m_{3/2} H_uH_d+h.c.\,,
\end{eqnarray}
which sets the scale of $B$ to be of order $m_{3/2}$ unless there is some tuning. However, there are additional contributions to $B$ because the $F$-terms of $P$ and $Q$ are non-zero giving the additional contributions
\begin{eqnarray}
-{\cal L}_{soft}\supset g \frac{F_QP+F_PQ}{\Lambda'} H_uH_d+h.c.
\end{eqnarray}
The $F$-terms for $P$ and $Q$ are
\begin{eqnarray}
F_Q=\lambda \frac{P^{\dagger 3}}{\Lambda}\,,  \\
F_P=3\lambda \frac{P^{\dagger 2}Q^\dagger}{\Lambda}\,.
\end{eqnarray}
If we use the relationships above for $P$ and $Q$ we find
\begin{eqnarray}
B=-\frac{81Z^8+(18y-6)Z^4+y(y-1)}{9Z^4+y}m_{3/2}\,.
\end{eqnarray}
If we now take $Z=1/2$ and $y=1$, as we did above, we find
\begin{eqnarray}
B=- .92 ~m_{3/2}\,.
\end{eqnarray}

Next we examine the contribution to the gaugino masses. The standard gauge mediation picture is to couple a pair of $\five$ and $\fivebar$'s to some spurion field
\begin{eqnarray}
W= \alpha Z \five\fivebar\,, \quad\quad \quad Z= M+\theta^2 F\,.
\end{eqnarray}
This interaction gives a contribution to the gaugino masses of
\begin{eqnarray}
\Delta M_i = \frac{g_i^2}{16\pi^2} \frac{F}{M}\,.
\end{eqnarray}
Since $P$ and $Q$ have non-zero $F$ terms, their couplings with pairs of $\five$ and $\fivebar$'s will induce just such a term.  In fact, we find a correction to the gaugino masses of
\begin{eqnarray}
\Delta M_i = \frac{g_i^2}{16\pi^2}\left(\frac{F_P}{\langle P \rangle} + 2 \frac{F_Q}{\langle Q \rangle }\right)\,.
\end{eqnarray}
The contributions can be parameterized in terms of $Z$,
\begin{eqnarray}
\frac{F_Q}{Q}=-\left(9Z^4+y\right)m_{3/2}=-\frac{25}{16}m_{3/2}\,,\label{FPFQ}\\
\frac{F_P}{P}=-\left(\frac{3Z^4}{9Z^4+y}\right)m_{3/2}=-\frac{3}{25}m_{3/2}\,, \nonumber
\end{eqnarray}
where the second equality is for $Z=1/2$ and $y=1$.  For this value of $Z$ and $y$, we get a contribution of
\begin{eqnarray}
\Delta M_i=-\frac{649}{200} \frac{g_i^2}{16\pi^2}m_{3/2}\,.
\end{eqnarray}
Since the change in the coefficients of the beta functions were increased by a factor of $3$ due to the additional $\five$ and $\fivebar$'s, this contribution completely cancel this contribution plus a little more. To leading order, the theory with this value of $Z$ would have gaugino masses very similar to the standard MSSM case. However, as can be seen from Eq. (\ref{FPFQ}), the gaugino masses scale with $Z^4$. So if we increase $Z$ a little the contribution to the gaugino masses will increase drastically.

\end{document}